%% file: paper.tex
\documentstyle[preprint,aps,eqsecnum,epsf,floats]{revtex}
\tighten

\begin{document}

\newcommand{\dencas}{\mbox{$m_\Lambda - m_\Xi$}}
\newcommand{\denn}{\mbox{$m_\Lambda - m_N$}}
\newcommand{\hf}{\mbox{$h_F$}}
\newcommand{\hd}{\mbox{$h_D$}}
\newcommand{\hc}{\mbox{$h_C$}}
\newcommand{\jay}[2]{\mbox{${\cal J}(\Delta m^{#1}_{#2})$}}
\newcommand{\h}{\mbox{${\cal H}$}}
\newcommand{\kay}[3]{\mbox{${\cal K}(#1,\Delta m^{#2}_{#3})$}}
\newcommand{\gee}[3]{\mbox{$G_{#1,#2,#3}$}}
\newcommand{\geetwid}[3]{\mbox{$\tilde G_{#1,#2,#3}$}}
\newcommand{\be}{\begin{eqnarray}}
\newcommand{\ee}{\end{eqnarray}}
\newcommand{\front}{\mbox{$
{m_K^2  \over 16\pi^2 f_K^2}\ln\left({m_K^2 \over\Lambda_\chi^2}\right)$}}
\newcommand{\llk}{\mbox{$\Lambda\Lambda K^0$}}
\newcommand{\hm}{\mbox{$\xi^\dagger h \xi$}}
\newcommand{\om}{\mbox{$\Omega^-$}}
\newcommand{\TTn}{\rule[-2.5mm]{0mm}{7.5mm}}

\def\cV{{\rm V}}
\def\cC{{\cal C}}
\def\cA{{\cal A}}
\def\cW{{\rm W}}
\def\si{{S_{int}}}
\def\cM{{\cal M}}
\def\Bp{B^+}
\def\BO{B^0}
\def\BsO{B_s^0}
\def\Pp{\pi^+}
\def\PO{\pi^0}
\def\Pm{\pi^-}
\def\KO{K^0}
\def\barKO{\overline K^0}
\def\Kp{K^+}
\def\Km{K^-}
\def\EA{\eta_8}
\def\EE{\eta_1}
\def\DO{D^0}
\def\Dp{D^+}
\def\Dsp{D_s^+}
\def\barDO{\overline D^0}
\def\Dm{D^-}
\def\Dsm{D_s^-}
\def\XbbO{\Xi_{bb}^0}
\def\Xbbm{\Xi_{bb}^-}
\def\Obbm{\Omega_{bb}^-}
\def\Xbcp{\Xi_{bc}^+}
\def\XbcO{\Xi_{bc}^0}
\def\ObcO{\Omega_{bc}^0}
\def\XbEm{\Xi_{b1}^-}
\def\XbEO{\Xi_{b1}^0}
\def\LbO{\Lambda_b^0}
\def\SbO{\Sigma_b^0}
\def\Sbm{\Sigma_b^-}
\def\Sbp{\Sigma_b^+}
\def\XbZO{\Xi_{b2}^0}
\def\XbZm{\Xi_{b2}^-}
\def\Obm{\Omega_b^-}
\def\JP{J/\Psi}
\def\Bm{B^-}
\def\barBO{\overline B^0}
\def\barBsO{\overline B_s^0}
\def\SO{\Sigma^0}
\def\LO{\Lambda^0}
\def\Sp{\Sigma^+}
\def\Sm{\Sigma^-}
\def\Xm{\Xi^-}
\def\XO{\Xi^0}
\def\pp{p}
\def\nn{n}
\def\II{I}
\def\DD{D}
\def\EE{E}
\def\NN{N}
\def\ALB{A_{\rm LB}}
\def\BLB{B_{\rm LB}}
\def\CLB{C_{\rm LB}}
\def\DLB{D_{\rm LB}}
\def\ELB{E_{\rm LB}}
\def\FLB{F_{\rm LB}}
\def\GLB{G_{\rm LB}}
\def\HLB{H_{\rm LB}}
\def\ILB{I_{\rm LB}}
\def\JLB{J_{\rm LB}}
\def\KLB{K_{\rm LB}}
\def\LLB{L_{\rm LB}}
\def\MLB{M_{\rm LB}}
\def\NLB{N_{\rm LB}}
\def\OLB{O_{\rm LB}}
\def\PLB{P_{\rm LB}}
\def\QLB{Q_{\rm LB}}
\def\RLB{R_{\rm LB}}
\def\SLB{S_{\rm LB}}
\def\TLB{T_{\rm LB}}
\def\ULB{U_{\rm LB}}
\def\VLB{V_{\rm LB}}
\def\WLB{W_{\rm LB}}
\def\XLB{X_{\rm LB}}
\newcommand{\TT}{\rule[-3mm]{0mm}{8mm}}

\def\DKS{[\frac{\partial}{\partial a} K_S]}
\def\DIm{[\frac{\partial}{\partial a} I_3]}

\def\scA{\cA_\si}
\def\scB{\cB_\si}
\def\dm{\Delta m}

%\preprint{PSI-PR-02-18;TUM/T39-02-21}
\preprint{\setlength{\baselineskip}{1.5em}
\small
\vbox{\vspace{-4cm}
\hbox{TUM/T39-02-21}
\hbox{PSI-PR-02-18}}}

\title{SU(3) Predictions for Weak Decays of Doubly Heavy Baryons -- including
SU(3) breaking terms}
\author{David A. Egolf}
\address{Department of Physics, Georgetown University, 37th and O St. NW,
Washington, D.C. 20057.
     \\  {\tt egolf@physics.georgetown.edu} }

\author{Roxanne P. Springer}
\address{Technische Universit\"at M\"unchen, Physik Department,
James-Franck-Strasse, 85748 Garching, Germany \footnote{Feb. 2002 - July 2002.}
 and Department of Physics, 
Duke University, Durham, NC 27708
\\ {\tt rps@phy.duke.edu}}
\author{J\"org Urban}
\address{Technische Universit\"at M\"unchen, Physik Department,
James-Franck-Strasse, 85748 Garching, Germany, and Paul Scherrer Institut,
CH-5232 Villigen PSI, Switzerland.
\\ {\tt joerg.urban@psi.ch}}
\maketitle
\begin{abstract}

We find expressions for the weak decay amplitudes of baryons
containing two $b$ quarks (or one $b$ and one $c$ quark -- many
relationship are the same) in terms of unknown reduced matrix
elements.  This project was originally motivated by the request of the
FNAL Run II $b$ Physics Workshop organizers for a guide to
experimentalists in their search for as yet unobserved hadrons.  We
include an analysis of linear SU(3) breaking terms in addition to
relationships generated by unbroken SU(3) symmetry, and relate these
to expressions in terms of the complete set of possible reduced matrix
elements.
\end{abstract}

%********************************************************************************
\section{Introduction}
\label{intro}
%********************************************************************************

The $b$ quark sector is currently of great interest for a number
of reasons.  First, the $b$ quark is the heaviest quark which lives long enough
to form hadrons.  Second, its unsuppressed coupling to the top quark
results in enhancements of a variety of decay modes.
Within standard
model physics, the $b$ quark sector provides an opportunity to 
probe the origin of $CP$ violation, and to judge whether it can
be accomodated entirely within the standard model.  Beyond the standard
model, the $b$ sector is sensitive to extra Higgs particles,
supersymmetric particles, and other types of new physics.
Finally, the current experimental situation with respect to $b$
physics is very active at both electron and hadron colliders.

Studying the weak decays of doubly heavy baryons allows us
to test our understanding both of different pieces of the
weak effective Lagrange density and of heavy--to--heavy and
heavy--to--light transitions \cite{quiggtalk}.  Understanding
the weak decay possibilities will aid in unravelling the
spectroscopy of doubly heavy baryons -- a rich field which
tests the applicability of Nonrelativistic QCD \cite{NRQCD}, 
Heavy Quark Effective Theory \cite{HQET}, 
sum rules \cite{sumrules}, and potential models
to these systems \cite{spec}.
On the experimental side, the first doubly heavy hadron containing
a $b$ quark was
seen at CDF in the form of $B_c$ \cite{bc}.  A doubly
charmed baryon may have been seen \cite{cchadron}.
Run IIb of the Tevatron at Fermilab is expected to produce 
doubly heavy hadrons at the nanobarn level \cite{workshop}.
Consideration of their spectroscopy has been considerable \cite{spec,workspec},
but only a few have looked at the baryon weak decay, especially for those
containing a $b$ quark \cite{wkdecay,onish}.  Here we study the weak
nonleptonic decay of doubly heavy baryons using SU(3) symmetry.

In order to identify new heavy
particles their weak decay modes should be analyzed to find the
most promising discovery modes and those modes which will give
the most physics insight.  To this end, we apply the 
procedure of imposing SU(3) symmetries and the Wigner-Eckart 
theorem for the decays of a number of particles of interest.
We also include the predictions which result from including linear
SU(3) breaking terms.  Finally, we point out how this analysis
relates to the general group theory basis for describing the
amplitudes.

This paper follows closely the procedures of, for example,
Refs.~\cite{quigg,sw1}
in identifying the relevant multiplets, 
forming the four-quark operators, 
decomposing them into irreducible SU(3)
representations, and building all possible singlets. 
Extensive studies of the $B$--meson and its decay modes have
been done in this manner \cite{Bmesons}, especially because of the potential 
for extracting CP--violating phases \cite{phases}. Baryons
containing a single $b$ quark have been studied this way in 
Ref.~\cite{bbaryons}, 
and those with two charmed quarks in Ref.~\cite{ccbaryon}.
 Our expressions
for decay amplitudes in terms of the most general group theoretic
basis is done following the procedure of Grinstein and Lebed, Ref.~\cite{gl}.
Tables of the results are presented,
and some of the relationships highlighted.  The Clebsch-Gordon
coefficients were found by writing a symbolic manipulation program 
which generated
the latex files of the tables appearing here.  The hope is that this
will substantially reduce the typographical errors which tend to occur in a
project such as this.

There are well-known examples of SU(3) violation in decays of
heavy particles.  The violation may come from the finite strange
quark mass, from final state interactions, or from nearby resonances.
SU(3) breaking has been studied in single--$b$ baryons in \cite{du},
and in $B$ mesons in \cite{Bbreak}, following treatments of lighter
systems \cite{savage,lightbreak}.  For the double heavy baryons, hints
of SU(3) violation are seen in potential model treatments \cite{kiselev}.
We hope that the analysis presented here will aid in the understanding of SU(3)
breaking effects in doubly heavy baryons, and 
illuminate which types fall with increasing
mass of the decaying particle.  

The paper is organized as follows:  in Section II the particles involved
in our weak decays are introduced and defined, and their
SU(3) transformation properties identified.  In Section III we discuss
the effective operators inducing the weak transitions.  We categorize
them based upon their degree of Cabbibo suppression, and then
decompose them into irreducible representations of SU(3).  In the
same section we include an SU(3) breaking effect for each operator
and decompose those terms under SU(3) as well. Details of the decomposition
are presented in Appendix~\ref{tendec}.  The decay amplitudes
themselves are found in Section IV, where we present tables of physical
amplitudes in terms of reduced matrix elements. Three very long tables
are relegated to Appendices \ref{3bb3bcMMa}, \ref{3bb3bcMMb}, and
\ref{3bb6bMtable}. We include in Section IV
a discussion of the relationship between the set of reduced amplitudes
found from including a linear SU(3) breaking term, and those found
when no assumptions are made. The details of how this is accomplished
is given in Appendix~\ref{matching}. The physical amplitudes expressed in
terms of the group theoretic basis is given in Appendix~\ref{completedecomp}. 
Rate relationships,
modulo phase space corrections, are also given in Section IV.  In 
Appendix~\ref{phasespacecorr} we 
discuss the effect of phase space corrections.  We conclude
with a summary in Section V.

%********************************************************************************
\section{Notation and Particle Multiplets}
\label{notation}
%********************************************************************************

We first consider decays of those particles identified as belonging
to the lowest mass SU(3) triplet we will call $3_{bb}$:
\begin{eqnarray}\label{3bb}
3_{bb} = \left( \Xi^0_{bb}, \ \Xi^-_{bb}, \ \Omega^-_{bb} 
    \right)^{\rm T} \ \ ,
\end{eqnarray}
with valence quarks ($bbu$, $bbd$, $bbs$).  The particle symbols are
dictated by their isospin quantum number.   These decay weakly
predominantly when one of the $b$ quarks decays.
The particle multiplets we will need for the final states are listed below.
\begin{enumerate}
\item The triplet of baryons with one $b$ and one $c$ quark are contained in
\begin{eqnarray}\label{3bc}
3_{bc} = \left( \Xi^+_{bc}, \ \Xi^0_{bc}, \ \Omega^0_{bc} 
    \right)^{\rm T}
\end{eqnarray}
with the quark content ($bcu$, $bcd$, $bcs$).
\item
The particles with one $b$ quark, which fall into the six representation 
of SU(3) and are therefore symmetric under interchange of indices, are:
\begin{eqnarray}\label{6b}
\ \ &
[6_b]^{11} = \Sigma^+_b \sim buu, & \ \ \ 
[6_b]^{12} = {1 \over \sqrt{2}}\Sigma^0_b \sim bud, \nonumber \\
&
[6_b]^{22} = \Sigma^-_b \sim bdd, & \ \ \ 
[6_b]^{13} = {1 \over \sqrt{2}}\Xi^0_{b2} \sim bus, \nonumber \\
&[6_b]^{33} = \Omega^-_b \sim bss, & \ \ \ 
[6_b]^{23} = {1 \over \sqrt{2}}\Xi^-_{b2} \sim bds.
\end{eqnarray}
\item The particles with one $b$ quark which fall into an
anti--triplet representation of SU(3) are:
\begin{eqnarray}\label{3b}
\overline 3_b = \left( \Xi^-_{b1}, \ - \Xi^0_{b1}, \ \Lambda^0_b
                \right),
\end{eqnarray}
with quark content ($bsd$, $bsu$, $bdu$).
\item The octet of lowest mass mesons are contained in
\begin{eqnarray}\label{octmes}
M = \left( \begin{array}{ccc}
{1\over \sqrt{2}}\pi^0  + {1\over \sqrt{6}}\eta_8  & \pi^+ & K^+ \\
\pi^- & -{1\over \sqrt{2}}\pi^0  + {1\over \sqrt{6}}\eta_8  & K^0 \\
K^- & \overline{K}^0 & -\sqrt{2 \over 3} \eta_8
\end{array}
\right) ,
\end{eqnarray}
with the quark content given in \cite{hg}.
\item The $D$--meson antitriplet is
\begin{eqnarray}\label{Dmes}
D =\left( D^0, \ D^+, \ D^+_s \right) ,
\end{eqnarray}
with valence quarks ($c \overline u$,  $c \overline d$,  $c \overline s$).
\item  $J/\Psi$ is a singlet under SU(3).
\item The $B$--meson antitriplet is
\begin{eqnarray}\label{Bmes}
B = \left( B^-, \ \overline B^0, \ \overline B_s^0
    \right) ,
\end{eqnarray}
with quark content ($b \overline u$, $b \overline d$, $b \overline
s$).
\item The lowest mass octet of baryons is
\begin{eqnarray}\label{boctet}
b =
\pmatrix{ {1\over\sqrt2}\Sigma^0 + {1\over\sqrt6}\Lambda^0 &
\Sigma^+ & p\cr \Sigma^-& -{1\over\sqrt2}\Sigma^0 +
{1\over\sqrt6}\Lambda^0&n\cr \Xi^- &\Xi^0 &- 
\sqrt{2\over 3}\Lambda^0\cr } ,
\end{eqnarray}
with quark content given in \cite{hg}.
\end{enumerate}
These multiplets will be used to build the decay amplitudes we will
analyze under SU(3).

%********************************************************************************
\section{The Effective Hamiltonian and Weak Four-Quark Operators}
\label{effham}
%********************************************************************************

We will group together decay processes which fall into the same
multiplets of initial and final states.  For instance, using
the notation above, the decay processes $3_{bb} \rightarrow
3_{bc} + M$ specify fourteen physical amplitudes.

As Grinstein and Lebed \cite{gl} point out, it is instructive
to re-write the decay amplitudes (the ``physical basis'') in terms of 
the ``group-theoretic'' basis where the amplitudes are written
in terms of reduced matrix elements of manifest SU(3) and isospin
content.  In the case $3_{bb} \rightarrow 3_{bc} + M$,
there are fourteen such reduced matrix elements since no assumptions
have been made about the Hamiltonian which induces the decay.
From this group-theoretic basis it is easier to see
how assumptions about the Hamiltonian (for instance, imposing SU(3) 
symmetry or including limited SU(3) breaking terms) can be used
to make predictions about relationships between decay amplitudes.
In Appendix \ref{completedecomp} we show this physical to group-theoretic basis
change for each of the processes we consider.  They are generated
using Mathematica \cite{wolfram}, using the necessary isoscalar
factors from Kaeding \cite{kaeding}.  Again, we follow the
prescription of Ref.~\cite{gl}.  

In order to obtain relationships between decay amplitudes within, for
instance, the $3_{bb} \rightarrow 3_{bc} + M$ processes, we make
assumptions about the Hamiltonian which induces the decay.  The
accuracy of these assumptions can then be tested experimentally.

The simplest choice is to match the Hamiltonian onto the four-quark
operators generated by the standard model weak Lagrangian:
\begin{eqnarray}\label{lag}
{\cal L}={g \over 2 \sqrt{2}} \left( \overline u \ \  \overline c \ \ 
    \overline t \right)
   \gamma^\mu (1-\gamma_5) \ V \ \left( \begin{array}{c}
                                             d \\ s \\ b \end{array}
    \right) \ W^+_\mu \ \ + \ \ {\rm h.c.}, 
\end{eqnarray}
where $V$ is the CKM matrix.  
We will use the Wolfenstein parameterization \cite{wolf} so that
\begin{eqnarray}\label{wolf1}
V \sim \left( \begin{array}{ccc}
1-\lambda^2/2 & \lambda & A \lambda^3 (\rho-i \eta) \\
-\lambda & 1- \lambda^2/2 & A \lambda^2 \\
A \lambda^3 (1-\rho-i \eta) & -A \lambda^2 & 1
\end{array}
\right) .
\end{eqnarray}
Following the procedure of \cite{sw1} we are interested in weak operators
with the following flavor quantum numbers:
\begin{enumerate}
\item $(b \overline c)(c \overline s)$ and  $(b \overline c)(c \overline d)$,
\item $(b \overline c)(u \overline s)$ and  $(b \overline c)(u \overline d)$,
\item $(b \overline u)(c \overline s)$ and  $(b \overline u)(c \overline d)$,
\item $(b \overline u)(u \overline s)$ and  $(b \overline u)(u \overline d)$.
\end{enumerate}

Since $(b \overline u)(c \overline d)$ and $(b \overline u)(u
\overline s)$ are highly Cabbibo-suppressed, we will neglect them.  We
decompose the remaining operators into irreducible SU(3) multiplets
which we will use, along with the multiplets of initial and final
states, to form expressions for the terms in our Hamiltonian relevant
to various decay processes.  Constants (the reduced matrix elements of
the Wigner-Eckart theorem) will remain undetermined in this
procedure.  Where an overall CKM parameter from the operators can be
absorbed into reduced matrix elements, we do so.  We give the SU(3)
decomposition of the four-quark operators above and then include the
effects of a term which transforms as $m_s (s \overline s)$, the
so-called ``linear breaking term,'' in an analysis similar to that in
Ref.~\cite{savage}.  This will allow us to see,
when data become available, if the dominant SU(3) breaking
comes from the strange quark mass.  Details of the decomposition,
including normalization, are given in Appendix \ref{tendec}.

\begin{enumerate}
\item 
The operators $(b \overline c) (c \overline s)$ and 
$(b \overline c) (c \overline d)$ transform as antitriplets and
\begin{eqnarray} \label{atrip}
(b \overline c) ( c \overline d) =
  H(\overline{3})_2 \sim V_{cb} \ V^*_{cd} ,
  \nonumber \\
(b \overline c) ( c \overline s) =  H(\overline{3})_3 \sim V_{cb} \ V^*_{cs} ,
  \nonumber \\
H_{(1)}(\overline{3}) \sim 
   \left( \begin{array}{c}
           0 \\
           -\lambda \\
           {\displaystyle{1-\frac{\lambda^2}{2}}}
    \end{array}\right)
\approx \left( \begin{array}{cc}
                     & 0 \\
                      -& \lambda \\
                       & 1 \end{array} \right) ,
\end{eqnarray}
where the overall CKM constant $A\lambda^2$ has been dropped since it will
not affect relationships between amplitudes induced by this operator.

Enlarging the group to incorporate the linear breaking terms
$(s \overline s) = 8 \oplus 1$, we have
$(b \overline c) (c \overline q) (s \overline s) \sim \overline{15} \oplus
6 \oplus \overline{3}_{(1)} \oplus \overline{3}_{(2)}$, with $\overline{q}=\overline{d}$ or
$\overline{s}$.
This gives the following nonzero elements in addition to the
$H_{(1)}(\overline 3)$ given above:

\begin{eqnarray}\label{atripss}
H(\overline{15})^3_{33} &=& 4 ,  \nonumber \\[2mm]
H(\overline{15})^2_{32} &=& 
H(\overline{15})^2_{23} = H(\overline{15})^1_{13} = 
H(\overline{15})^1_{31} = -2 ,  \nonumber \\[2mm]
H(\overline{15})^3_{23} &=& 
H(\overline{15})^3_{32}= -3 \ \lambda  ,  \nonumber \\[2mm]
H(\overline{15})^1_{21} &=& 
H(\overline{15})^1_{12}= \lambda  ,  \nonumber \\[2mm]
H(\overline{15})^2_{22} &=& 2 \ \lambda ,  \nonumber \\[2mm]
H(6)^3_{23}&=&
H(6)^1_{12}= -H(6)^1_{21}= -H(6)^3_{32}= -\lambda , \nonumber \\[2mm]
H_{(2)}(\bar{3})_2 &=& \lambda, \nonumber \\[2mm]
H_{(2)}(\bar{3})_3 &=& 2,
\end{eqnarray}
where again we keep only the CKM parameter $\lambda$ explicit.
The elements of the 6 are given in terms of their Levi-Civita tensor
contracted form.

\item The operators $(b \overline c) ( u \overline d)$ and
$(b \overline c) ( u \overline s)$ transform as octets under SU(3):

\begin{eqnarray} \label{oct}
(b \overline c) ( u \overline d) = H_{(1)}(8)^1_2 \sim V_{cb} \ V^*_{ud} ,
\nonumber \\
(b \overline c) ( u \overline s) = H_{(1)}(8)^1_3 \sim V_{cb} \ V^*_{us} ,
\nonumber \\
H_{(1)}(8)
  \sim \left( \begin{array}{ccc}
               0&\;\;\;1-{\displaystyle\frac{\lambda^2}{2}}\;\;\;& \lambda \\
               0&0&0\\
               0&0&0 \end{array} \right)
  \approx \left( \begin{array}{ccc}
               0&1& \lambda \\
               0&0&0\\
               0&0&0 \end{array} \right) \ \ .
\end{eqnarray}

Including the linear breaking term we have
\begin{eqnarray}
(b \overline c) ( u \overline q) (s \overline s) \sim 8_{(1)} \otimes (8\oplus 1) = 
27 \oplus 10 \oplus \overline{10} \oplus 8_{(1)} \oplus 8_{(2)} \oplus 8_{(3)},
\end{eqnarray} 
where $\overline q=\overline d$ or $\overline s$.  $8_{(3)}$ is
already saturated by $8_{(1)}$ for our purposes, and furthermore there is no
singlet for this particular operator. 
The nonzero operator terms in addition to those in Eqn.~\ref{oct} above are:
\begin{eqnarray}\label{octss}
H(27)^{13}_{23} &=& 
H(27)^{31}_{23} = 
H(27)^{13}_{32} = 
H(27)^{31}_{32} = 2 ,  \nonumber \\[2mm]
H(27)^{12}_{22} &=& 
H(27)^{21}_{22}  = 
H(27)^{11}_{21} = 
H(27)^{11}_{12}= -1 ,  \nonumber \\[2mm]
H(27)^{12}_{23}&=&
H(27)^{12}_{32}=
H(27)^{21}_{23}=
H(27)^{21}_{32}= -\lambda ,  \nonumber \\[2mm]
H(27)^{11}_{13}&=&
H(27)^{11}_{31}= -2 \ \lambda ,  \nonumber \\[2mm]
H(27)^{13}_{33}&=&
H(27)^{31}_{33}= 3 \ \lambda  ,  \nonumber \\[2mm]
H(10)^{13}_{23}&=&
-H(10)^{31}_{32}=
-H(10)^{13}_{32}=
H(10)^{31}_{23}=1,  \nonumber \\[2mm]
H(10)^{11}_{12} &=& 
-H(10)^{11}_{21}=1  ,  \nonumber \\[2mm]
H(\overline{10})^{13}_{23}&=&
-H(\overline{10})^{31}_{23}=
-H(\overline{10})^{31}_{32}=
H(\overline{10})^{13}_{32}=1  ,  \nonumber \\[2mm]
H(\overline{10})^{21}_{22} &=& 
-H(\overline{10})^{12}_{22}=1  ,  \nonumber \\[2mm]
H(\overline{10})^{21}_{32}&=&
H(\overline{10})^{21}_{23}=
-H(\overline{10})^{12}_{23}=
-H(\overline{10})^{12}_{32}=\lambda ,  \nonumber \\[2mm]
H(\overline{10})^{13}_{33} &=& 
-H(\overline{10})^{31}_{33}=\lambda ,  \nonumber \\[2mm]
H_{(2)}(8)^1_2 &=& -1, \nonumber \\[2mm]
H_{(2)}(8)^1_3 &=& 2 \ \lambda, 
\end{eqnarray}
where the elements of the $10$ and $\overline{10}$ are given in terms
of their Levi-Civita tensor contracted form.  

\item The operator $(b \overline u)(c \overline s)$ decomposes into a 3 and
a $\overline 6$, with nonzero elements
\begin{eqnarray}\label{bucs}
H^\prime(\overline 6)_{13}&=&
H^\prime(\overline 6)_{31}= 1 ,  \nonumber \\[2mm]
H^\prime(3)^2&=&1 ,
\end{eqnarray}
where we now use $H^\prime$ to distinguish these operator elements
from the ones appearing previously to guard against possible
confusion. No overall CKM constants are necessary because this
operator will not appear in the same process with any
other operators.

Including a linear breaking term gives $(b \overline u)(c \overline s)
(s \overline s) \sim \overline{24} \oplus 15 \oplus \overline 6 \oplus 3$
with nonzero elements:
\begin{eqnarray}
\label{bucslin}
H^\prime(\overline{24})^2_{321}&=&
H^\prime(\overline{24})^2_{231}=
H^\prime(\overline{24})^2_{213}=
H^\prime(\overline{24})^2_{123}=
H^\prime(\overline{24})^2_{312}=
H^\prime(\overline{24})^2_{132}=-1
,  \nonumber \\[2mm]
H^\prime(\overline{24})^1_{113}&=&
H^\prime(\overline{24})^1_{131}=
H^\prime(\overline{24})^1_{311}=-2
,  \nonumber \\[2mm]
H^\prime(\overline{24})^3_{331}&=&
H^\prime(\overline{24})^3_{133}=
H^\prime(\overline{24})^3_{313}= 3 
,  \nonumber \\[2mm]
H^\prime(15)^{32}_3&=&
H^\prime(15)^{23}_3 = 3 
,  \nonumber \\[2mm]
H^\prime(15)^{21}_1 &=& 
H^\prime(15)^{12}_1 = -1
,  \nonumber \\[2mm]
H^\prime(15)^{22}_2 &=& -2,
\end{eqnarray}
and the $\overline{6}$ and 3 the same as in Eqn.~\ref{bucs} above.

\item Finally we have the decomposition of the operator
$(b \overline u)(u \overline d)$ into irreducible SU(3)
representations.  This operator induces the same decays as those in
Eqn.~\ref{atrip} and so they will appear together in
amplitude expressions. Therefore we need to include the overall
CKM factor  $\lambda (\rho - i \eta)$.  In the elements listed
below we neglect CP violation and use only $\lambda \rho$;
the $\eta$ dependence can be recapured by substitution.
We have
\begin{eqnarray}
(b \overline u) (u \overline d) \sim \overline{15}_{(1)}
\oplus 6 \oplus \overline{3},
\end{eqnarray}
where 
\begin{eqnarray}\label{light}
H_{(1)}^{\prime \prime}(\overline{15})_{12}^1 &=& 
H_{(1)}^{\prime \prime}(\overline{15})_{21}^1 = 3 \ \lambda \ \rho
,  \nonumber \\[2mm]
H_{(1)}^{\prime \prime}(\overline{15})_{22}^2 &=& -2 \ \lambda \ \rho
,  \nonumber \\[2mm]
H_{(1)}^{\prime \prime}(\overline{15})_{32}^3 &=& 
H_{(1)}^{\prime \prime}(\overline{15})_{23}^3 = - \lambda \ \rho ,  \nonumber \\[2mm]
H^{\prime \prime}(6)^{13} &=& 
H^{\prime \prime}(6)^{31} = \lambda \ \rho ,  \nonumber \\[2mm]
H^{\prime \prime} (\overline 3)_2&=& \lambda \ \rho.
\end{eqnarray}
We will also need the 6 in its Levi-Civita tensor contracted form:
$H^{\prime \prime}(6)_{12}^1 = - 
H^{\prime \prime}(6)_{21}^1 =  \lambda \ \rho
\ ; \
H^{\prime \prime}(6)_{23}^3 = - H^{\prime \prime}(6)_{32}^3 = \lambda \ \rho $.

Including the linear breaking term gives 
\begin{eqnarray}
(b \overline u) (u \overline d) (s \overline s) 
&\sim& (\overline{15}_{(1)}\oplus 6 \oplus \overline{3}) \otimes (8 \oplus 1) 
\nonumber\\&=&
\overline{42} \oplus 24_{(1)} \oplus 24_{(2)} \oplus \overline{15'} 
\oplus \overline{15}_{(1)} \oplus \overline{15}_{(2)} \oplus \overline{15}_{(3)} 
\oplus \overline{15}_{(4)} \oplus 6 \oplus \overline{3}.
\end{eqnarray}
There is no additional $6$ or $\bar{3}$ compared to Eqn. \ref{light}.
The $\overline{15}_{(1)}$, the 6, and the $\overline{3}$ are given above,
while the remaining nonzero elements are:\\[3mm]
i) from the tensor product $\bar{3} \otimes 8$:
\begin{eqnarray}\label{lightss3}
H_{(2)}^{\prime \prime}(\overline{15})_{12}^1 &=& 
H_{(2)}^{\prime \prime}(\overline{15})_{21}^1 = -\lambda \ \rho
,  \nonumber \\[2mm]
H_{(2)}^{\prime \prime}(\overline{15})_{22}^2 &=& -2 \ \lambda \ \rho
,  \nonumber \\[2mm]
H_{(2)}^{\prime \prime}(\overline{15})_{32}^3 &=& 
H_{(2)}^{\prime \prime}(\overline{15})_{23}^3 = 3 \ \lambda \ \rho,
\end{eqnarray}
ii) from the tensor product $6 \otimes 8$:\nopagebreak
\begin{eqnarray}\label{lightss6}
H^{\prime \prime}_{(1)}(24)^{321}_2 &=&
H^{\prime \prime}_{(1)}(24)^{231}_2=
H^{\prime \prime}_{(1)}(24)^{213}_2= - \lambda \ \rho ,  \nonumber \\[2mm]
H^{\prime \prime}_{(1)}(24)^{123}_2 &=&
H^{\prime \prime}_{(1)}(24)^{312}_2=
H^{\prime \prime}_{(1)}(24)^{132}_2= - \lambda \ \rho ,  \nonumber \\[2mm]
H^{\prime \prime}_{(1)}(24)^{113}_1 &=&
H^{\prime \prime}_{(1)}(24)^{131}_1=
H^{\prime \prime}_{(1)}(24)^{311}_1= - 2 \ \lambda \ \rho ,  \nonumber \\[2mm]
H^{\prime \prime}_{(1)}(24)^{331}_3 &=&
H^{\prime \prime}_{(1)}(24)^{133}_3=
H^{\prime \prime}_{(1)}(24)^{313}_3= 3 \ \lambda \ \rho,
\end{eqnarray}
iii) from the tensor product $\overline{15}_{(1)} \otimes 8$:
\begin{eqnarray}\label{lightss15}
H^{\prime \prime}(\overline{42})^{11}_{121} &=& 
H^{\prime \prime}(\overline{42})^{11}_{112} = 
H^{\prime \prime}(\overline{42})^{11}_{211} = 
H^{\prime \prime}(\overline{42})^{33}_{323} =
H^{\prime \prime}(\overline{42})^{33}_{332} = 
H^{\prime \prime}(\overline{42})^{33}_{233} = 
8\ \lambda \ \rho ,  \nonumber \\[2mm]
H^{\prime \prime}(\overline{42})^{22}_{222} &=& 
-6\ \lambda \ \rho ,  \nonumber \\[2mm]
H^{\prime \prime}(\overline{42})^{12}_{221} &=& 
H^{\prime \prime}(\overline{42})^{12}_{212} = 
H^{\prime \prime}(\overline{42})^{12}_{122} = 
H^{\prime \prime}(\overline{42})^{21}_{221} = 
H^{\prime \prime}(\overline{42})^{21}_{212} = 
H^{\prime \prime}(\overline{42})^{21}_{122} = 
3  \ \lambda \ \rho ,  \nonumber \\[2mm]
H^{\prime \prime}(\overline{42})^{23}_{322} &=& 
H^{\prime \prime}(\overline{42})^{23}_{232} = 
H^{\prime \prime}(\overline{42})^{23}_{223} = 
H^{\prime \prime}(\overline{42})^{32}_{322} = 
H^{\prime \prime}(\overline{42})^{32}_{232} =
H^{\prime \prime}(\overline{42})^{32}_{223} = 
3  \ \lambda \ \rho,\nonumber\\[2mm]
H^{\prime \prime}(\overline{42})^{13}_{132} &=& 
H^{\prime \prime}(\overline{42})^{13}_{123} = 
H^{\prime \prime}(\overline{42})^{13}_{231} = 
H^{\prime \prime}(\overline{42})^{13}_{213} = 
-11  \ \lambda \ \rho  ,  \nonumber \\[2mm]
H^{\prime \prime}(\overline{42})^{13}_{312} &=& 
H^{\prime \prime}(\overline{42})^{13}_{321} =
H^{\prime \prime}(\overline{42})^{31}_{132} = 
H^{\prime \prime}(\overline{42})^{31}_{123} =
-11  \ \lambda \ \rho  ,  \nonumber \\[2mm]
H^{\prime \prime}(\overline{42})^{31}_{231} &=& 
H^{\prime \prime}(\overline{42})^{31}_{213} =
H^{\prime \prime}(\overline{42})^{31}_{312} = 
H^{\prime \prime}(\overline{42})^{31}_{321} = 
-11 \ \lambda \ \rho ,  \nonumber \\[2mm]
H^{\prime \prime}_{(2)}(24)^{132}_2 &=&
H^{\prime \prime}_{(2)}(24)^{123}_2=
H^{\prime \prime}_{(2)}(24)^{213}_2= - 3 \ \lambda \ \rho ,  \nonumber \\[2mm]
H^{\prime \prime}_{(2)}(24)^{231}_2 &=&
H^{\prime \prime}_{(2)}(24)^{312}_2=
H^{\prime \prime}_{(2)}(24)^{321}_2= - 3 \ \lambda \ \rho ,  \nonumber \\[2mm]
H^{\prime \prime}_{(2)}(24)^{113}_1 &=&
H^{\prime \prime}_{(2)}(24)^{131}_1=
H^{\prime \prime}_{(2)}(24)^{311}_1= 4 \ \lambda \ \rho ,  \nonumber \\[2mm]
H^{\prime \prime}_{(2)}(24)^{331}_3 &=&
H^{\prime \prime}_{(2)}(24)^{133}_3=
H^{\prime \prime}_{(2)}(24)^{313}_3= -\lambda \ \rho, \nonumber \\[2mm]
H_{(3)}^{\prime \prime}(\overline{15})_{12}^1 &=& 
H_{(3)}^{\prime \prime}(\overline{15})_{21}^1 = -9 \ \lambda \ \rho
,  \nonumber \\[2mm]
H_{(3)}^{\prime \prime}(\overline{15})_{22}^2 &=& 14 \ \lambda \ \rho
,  \nonumber \\[2mm]
H_{(3)}^{\prime \prime}(\overline{15})_{32}^3 &=& 
H_{(3)}^{\prime \prime}(\overline{15})_{23}^3 = - 5 \ \lambda \ \rho , \nonumber \\[2mm]
H_{(4)}^{\prime \prime}(\overline{15})_{12}^1 &=& 
H_{(4)}^{\prime \prime}(\overline{15})_{21}^1 = -21 \ \lambda \ \rho
,  \nonumber \\[2mm]
H_{(4)}^{\prime \prime}(\overline{15})_{22}^2 &=& 22 \ \lambda \ \rho
,  \nonumber \\[2mm]
H_{(4)}^{\prime \prime}(\overline{15})_{32}^3 &=& 
H_{(4)}^{\prime \prime}(\overline{15})_{23}^3 = - \lambda \ \rho ,
\nonumber \\[2mm]
H^{\prime \prime}(\overline{15}^\prime)_{2321} &=& \lambda \ \rho ,
\label{ham4part3}
\end{eqnarray} 
along with the other 11 components which come from the fact that
$H^{\prime \prime}(\overline{15}^\prime)$ is totally symmetric
over its four indices.\\[3mm]
\end{enumerate}
For our purposes, the $H_{(1)}^{\prime \prime}(\overline{15})$
and $H_{(2)}^{\prime \prime}(\overline{15})$ are sufficient to provide
a minimal set of unknowns, so we do not use 
$\overline{15}_{(3)}$ or $\overline{15}_{(4)}$.
That is, the reduced amplitudes associated with $\overline{15}_{(3)}$ 
and $\overline{15}_{(4)}$ are linear combinations of the
reduced amplitudes associated with $\overline{15}_{(1)}$ and
 $\overline{15}_{(2)}$.

Notice, as suggested in Ref.~\cite{gl}, that this linear breaking
term actually saturates more general SU(3) breaking than simply
$(s \overline s)$.  If, for instance, we chose to include a term
$(u \overline u)$ in extending the operators from Eqn.~\ref{atrip},
we would find that it is already mimicked in Eqn.~\ref{light}
(albeit with some $\rho$ dependence now absorbed into the unknown 
constants).  
However, including the linear breaking $(s \overline s)$ term
does not saturate arbitrary SU(3) breaking, which is why we are
still left with predictions.

%********************************************************************************
\section{Decay Amplitudes}
\label{decamps}
%********************************************************************************
%
%********************************************************
\subsection{$3_{bb} \rightarrow 3_{bc} \ + \ M$ (
Final states with a b=--1, c=1 triplet baryon plus an octet meson)}
%********************************************************

This is the first of the decay amplitudes induced by Eqn.~\ref{oct}.
From the product of $3_{bb}$, $3_{bc}$, $M$, and the operators
in Eqn.~\ref{oct} we can form
three independent singlets, and so have expressions for all fourteen
physical amplitudes in terms of only three unknown reduced matrix
elements. We
call them $A$, $B$, $C$, etc.  We will use this notation for each
type of decay, even though their values for one type of process
are unrelated to those of another.  Since the equations and tables
will occur in different sections of the paper, this should not
cause confusion.

Including the extension to linear breaking embodied in
Eqn.~\ref{octss},
we can form six more singlets with this additional freedom.
We call the unknown constants associated with each of these
$A_{LB}$, $B_{LB}$, etc., where $LB$ stands for ``linear breaking''
even though, as we have seen, it can be more general than this.
With this looser assumption about the Hamiltonian, we have
fourteen amplitudes expressed in terms of nine reduced matrix elements.
The expressions are listed in Table~\ref{3bcM}.  In order
to recapture the amplitudes in the limit of exact SU(3) symmetry,
we can set all of the $LB$ coefficients to zero.

\begin{eqnarray}
{\cal H}_{LB}(3_{bb} \rightarrow 3_{bc} \ + \ M) &=&
   A \ [\overline 3_{bb}]_i \ [3_{bc}]^i M^k_j H_{(1)}(8)^j_k 
   + B \ [\overline 3_{bb}]_i \ [3_{bc}]^j M^i_k H_{(1)}(8)^k_j \nonumber \\
&+& C \ [\overline 3_{bb}]_i \ [3_{bc}]^j M^k_j H_{(1)}(8)^i_k \nonumber \\
&+& A_{LB} \ [\overline 3_{bb}]_i \ [3_{bc}]^i M^k_j H_{(2)}(8)^j_k
   + B_{LB} \ [\overline 3_{bb}]_i \ [3_{bc}]^j M^i_k H_{(2)}(8)^k_j \nonumber \\
&+& C_{LB} \ [\overline 3_{bb}]_i \ [3_{bc}]^j M^k_j H_{(2)}(8)^i_k
   + D_{LB} \ 
    [\overline 3_{bb}]_i \ [3_{bc}]^k M^l_j H(10)^{ij}_{kl} \nonumber \\
&+& E_{LB} \ 
    [\overline 3_{bb}]_i \ [3_{bc}]^k M^l_j H(\overline{10})^{ij}_{kl}
   + F_{LB} \ 
    [\overline 3_{bb}]_i \ [3_{bc}]^k M^l_j H(27)^{ij}_{kl}.
\end{eqnarray}

\begin{table}[!thb]
\begin{center}
\mbox{\epsffile{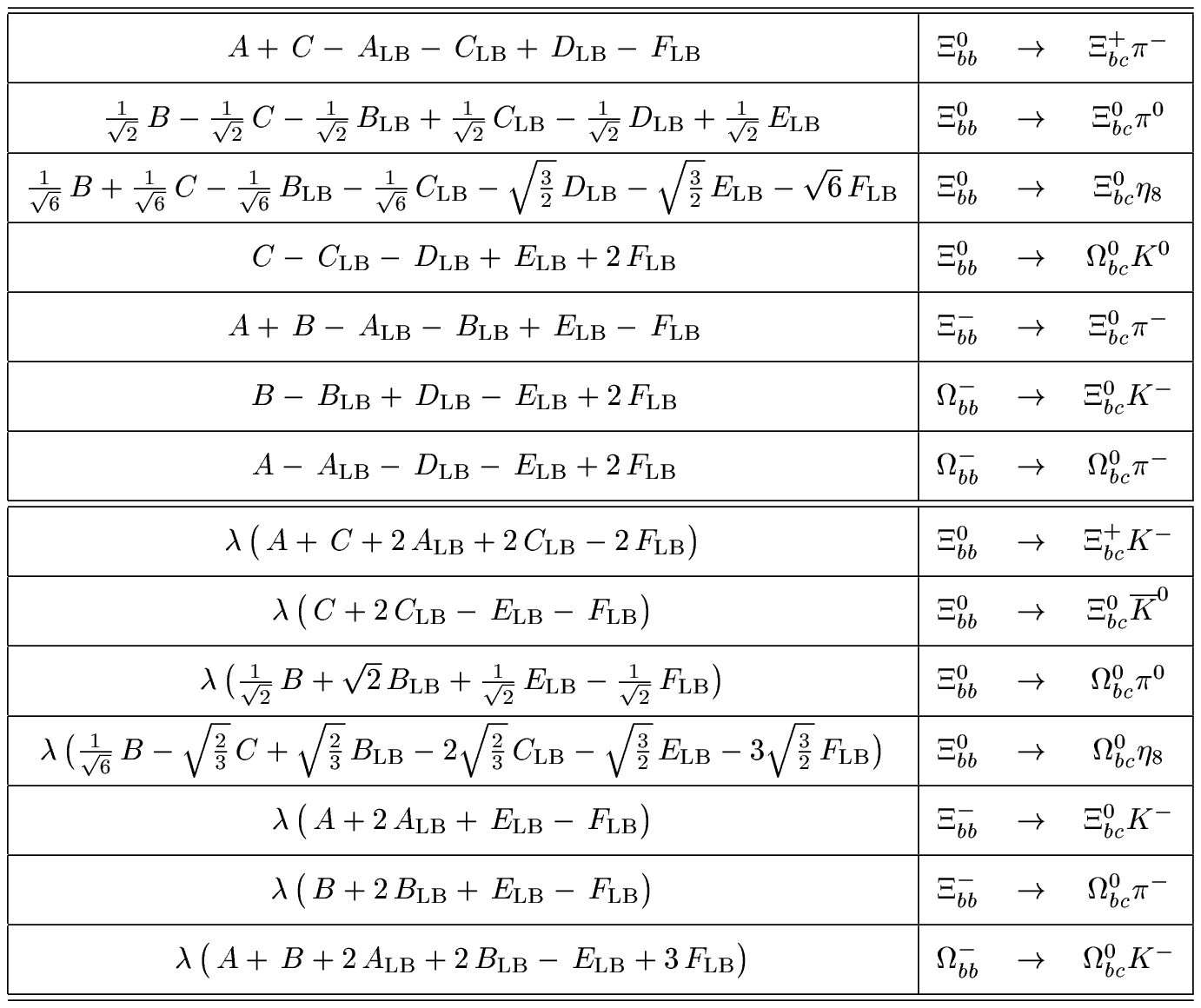}}
\caption{Matrix elements for the decay $3_{bb} \rightarrow 3_{bc} +  M$ \label{3bcM}}
\end{center}
\end{table}

Examples
of relationships between Cabibbo suppressed and Cabibbo allowed decays
which occur under exact SU(3) are (see table \ref{3bcM})\footnote{In writing
these rate relationships we are ignoring potential phase space issues.
See Appendix~\ref{phasespacecorr} for a discussion.}:
\begin{eqnarray}\label{bcMrate}
\Gamma\left(\Xi^0_{bb} \rightarrow \Xi^{+}_{bc} \ \pi^{-}\right)
 &=& {1 \over \lambda^2} \
  \Gamma\left(\Xi^0_{bb} \rightarrow \Xi^{+}_{bc} \ K^{-}\right) , \nonumber
\\
\Gamma\left(\Xi^{-}_{bb} \rightarrow \Xi^0_{bc} \ \pi^{-}\right)
 &=& {1 \over \lambda^2} \
  \Gamma\left(\Omega^{-}_{bb} \rightarrow \Omega^{0}_{bc} \ K^{-}\right).
\end{eqnarray}
From this and the branching ratio for the decay $\Xi^{0/-}_{bb} 
\rightarrow
\Xi^{+/0}_{bc} \pi^-$ found using potential models ($2.2 \%$)
\cite{onish}, we
would predict that 
${\rm BR}\left(\Xi^0_{bb} \rightarrow \Xi^{+}_{bc} \ K^{-}\right) \sim
0.1 \%$.
The extent to which such relationships do not hold experimentally is
a measure of SU(3) symmetry breaking in these decays.  Semileptonic
decays can be treated in the same way. For $3_{bb} \rightarrow
3_{bc} \ l \ \overline \nu$ the results are trivial unless SU(3)
breaking is included, in which case only isospin relationships survive.

The reduced matrix elements induced by arbitrary SU(3) breaking,
and their relationships
to the fourteen allowed processes in this category, are given in
Appendix~\ref{3bb3bcM}.  
Matching these results to those where only linear breaking of SU(3) 
is imposed (Table \ref{3bcM}), 
according to the description in Appendix~\ref{matching}, 
we find the following relationships for these matrix elements:
\begin{eqnarray}\label{rel1}
\langle \bar{6}||\overline{10}_{I=\frac{3}{2}} ||3\rangle &=& 0,\nonumber\\
\langle 15||27_{I=\frac{3}{2}} ||3\rangle &=& 0,\nonumber\\
\langle 15||27_{I=2} ||3\rangle &=& 0,\nonumber\\
\langle 15||10_{I=\frac{1}{2}}||3 \rangle &=& \lambda \ \langle 15 ||10_{I=1}||3 \rangle
  , \nonumber\\
\langle 15||27_{I=\frac{1}{2}}||3 \rangle &=& \sqrt{3 \over 2} \ \lambda \
  \langle 15 ||27_{I=1}||3 \rangle .
\end{eqnarray}
It is clear why the higher isospin reduced amplitudes are zero for
linear breaking:  extension by $(s \overline s)$ cannot change the
isospin of the underlying octet operators in Eqn.~\ref{oct}, 
which have $I=1/2$ for
the Cabbibo suppressed decay and $I=1$ for the Cabbibo allowed decay.

Note that if the $b$ quark decays first, a table with entries
identical to Table~\ref{3bcM} applies to the analogous decay $3_{bc}
\rightarrow 3_{cc} + M$, where $3_{cc}$ are the triplet of doubly
charmed baryons with quark quantum numbers ($ccu,ccd,ccs$).  This is
obtained by making the $b \rightarrow c$ substitution in the
right-hand column (with the appropriate changes in charge and particle
symbol).  Pauli suppression factors are the same for each decay and so
do not affect relationships among them.

%********************************************************
\subsection{$3_{bb} \rightarrow 3_{bc} \ + \ M \ + \ M$
(Final states with a b=--1, c=1 triplet 
baryon plus two octet mesons)}
%********************************************************

The operators which induce the $3_{bb}$ decay into $3_{bc}$ plus two
mesons are the same as those in the last section.  Now however we have
many more possible ways of creating a singlet using these tensors,
resulting in a larger set of reduced matrix elements in which to
express decay rates.  Further, because we are dealing with two meson octets
in the final state -- which are identical under SU(3) -- we must impose
the appropriate symmetry.  For mesons in a relative even angular momentum
state $L$, our Hamiltonian is symmetric under interchange of SU(3) indices
on the meson tensors.  For mesons in a relative odd angular momentum
state $L$, our Hamiltonian is antisymmetric under interchange of SU(3)
indices on the meson tensors.  Forty-six even--$L$ physical decays 
are expressed in
terms of six reduced matrix elements when SU(3) symmetry is imposed.
There are forty-two odd--$L$ physical amplitudes, where because
of the symmetry requirement there are only five independent reduced
matrix elements.  Because of the large
number of additional reduced matrix elements appearing when the SU(3)
symmetry is broken by a linear term, we do not present tables including
them.  The SU(3) conserving predictions are found in Appendices
\ref{3bb3bcMMa} and \ref{3bb3bcMMb}.  These tables list the decay
``amplitudes'' by already taking into account the symmetry factors
for identical mesons in the final state.  So, excepting mass corrections
in the phase space, 
rates are obtained simply by taking the absolute value squared
of the expressions given in Appendices \ref{3bb3bcMMa} and \ref{3bb3bcMMb}.
Note that no identical mesons are found in the final state of an $L$--odd
decay because of symmetry restrictions.  However, two final states occur
for $L$--odd which do not appear for $L$--even:  
$\Omega_{bb}^- \rightarrow \Omega_{bc}^0 \pi^0 \pi^-$ for the 
Cabibbo allowed case, and
$\Xi_{bb}^- \rightarrow \Omega_{bc}^0 \pi^0 \pi^-$ for the Cabibbo
suppressed case.  For the $L$--even case we have:

\begin{eqnarray}
{\cal H}_{LB}(3_{bb} \rightarrow 3_{bc} + M + M)&=&\nonumber \\
&&\hspace{-3cm}+A \ [\overline 3_{bb}]_i \ [3_{bc}]^i M^j_k M^k_l H_{(1)}(8)^l_j
   +B \ [\overline 3_{bb}]_i \ [3_{bc}]^j M^i_j M^l_k H_{(1)}(8)^k_l \nonumber \\
&&\hspace{-3cm}+C \ [\overline 3_{bb}]_i \ [3_{bc}]^j M^i_k M^l_j H_{(1)}(8)^k_l
   +D \ [\overline 3_{bb}]_i \ [3_{bc}]^j M^i_l M^l_k H_{(1)}(8)^k_j \nonumber \\
&&\hspace{-3cm}+E \ [\overline 3_{bb}]_i \ [3_{bc}]^j M^k_j M^l_k H_{(1)}(8)^i_l
  +F \ [\overline 3_{bb}]_i \ [3_{bc}]^j M^l_k M^k_l H_{(1)}(8)^i_j \nonumber \\
&&\hspace{-3cm}+A_{LB} \ [\overline 3_{bb}]_i \ [3_{bc}]^i M^j_k M^k_l H_{(2)}(8)^l_j
  +B_{LB} \ [\overline 3_{bb}]_i \ [3_{bc}]^j M^i_j M^l_k H_{(2)}(8)^k_l \nonumber \\
&&\hspace{-3cm}+C_{LB} \ [\overline 3_{bb}]_i \ [3_{bc}]^j M^i_k M^l_j H_{(2)}(8)^k_l
  +D_{LB} \ [\overline 3_{bb}]_i \ [3_{bc}]^j M^i_l M^l_k H_{(2)}(8)^k_j \nonumber \\
&&\hspace{-3cm}+E_{LB} \ [\overline 3_{bb}]_i \ [3_{bc}]^j M^k_j M^l_k H_{(2)}(8)^i_l
  +F_{LB} \ [\overline 3_{bb}]_i \ [3_{bc}]^j M^l_k M^k_l H_{(2)}(8)^i_j \nonumber \\
&&\hspace{-3cm}+G_{LB} \ [\overline 3_{bb}]_i \ [3_{bc}]^j M^i_p M^k_q H(10)^{pq}_{jk} 
  +H_{LB} \ [\overline 3_{bb}]_i \ [3_{bc}]^j M^k_j M^r_q H(10)^{iq}_{kr} 
  \nonumber \\
&&\hspace{-3cm}+I_{LB} \ [\overline 3_{bb}]_i \ [3_{bc}]^j M^k_q M^q_r H(10)^{ir}_{jk} 
  +J_{LB} \ [\overline 3_{bb}]_i \ [3_{bc}]^j M^i_p M^k_q 
  H(\overline{10})^{pq}_{jk} 
\nonumber \\
&&\hspace{-3cm}
  +K_{LB} \ [\overline 3_{bb}]_i \ [3_{bc}]^j M^k_j M^r_q 
  H(\overline{10})^{iq}_{kr} 
 +L_{LB} \ [\overline 3_{bb}]_i \ [3_{bc}]^j M^k_q M^q_r 
  H(\overline{10})^{ir}_{jk} \nonumber \\
&&\hspace{-3cm}+M_{LB} \ [\overline 3_{bb}]_i \ [3_{bc}]^i M^j_r M^k_p H(27)^{rp}_{jk} 
  +N_{LB} \ [\overline 3_{bb}]_i \ [3_{bc}]^j M^i_p M^k_q H(27)^{pq}_{jk} 
  \nonumber \\
&&\hspace{-3cm}+O_{LB} \ [\overline 3_{bb}]_i \ [3_{bc}]^j M^k_j M^r_q H(27)^{iq}_{kr} 
  +P_{LB} \ [\overline 3_{bb}]_i \ [3_{bc}]^j M^k_q M^q_r H(27)^{ir}_{jk}.
\end{eqnarray}

For the
unbroken SU(3) case there are many relationships. Below are some
examples.  If there is no subscript on the meson states then the
relationship holds independently of whether $L$ is even or odd.
\begin{eqnarray}
2 \ \Gamma\Big(\Xi^0_{bb} \rightarrow \Xi^+_{bc} \  
       (\pi^0 \ \pi^-)_{L=even}\Big) &=&
{1 \over 2} \ 
     \Gamma\Big(\Xi^-_{bb} \rightarrow \Xi^+_{bc} \ (\pi^- \ \pi^-)_{L=even}
  \Big) =
  \nonumber \\
2 \ \Gamma\Big(\Xi^-_{bb} \rightarrow \Xi^0_{bc} \ (\pi^0 \ \pi^-)_{L=even}
  \Big) &=&
\Gamma\Big(\Xi^-_{bb} \rightarrow \Omega^0_{bc} \ (\pi^- \ K^0)_{L=even}
  \Big) =
  \nonumber \\
\Gamma\Big(\Omega^-_{bb} \rightarrow \Xi^+_{bc} \ (\pi^- \ K^-)_{L=even}
  \Big)
 & =& {1 \over \lambda^2} \ 
   \Gamma\Big(\Xi^-_{bb} \rightarrow \Xi^+_{bc} \ (\pi^- \ K^-)_{L=even}
  \Big) =
\nonumber \\
 {1 \over \lambda^2} \
   \Gamma \Big(\Omega^-_{bb} \rightarrow \Xi^0_{bc} \ 
                       (K^- \ \overline K^0)_{L=even}\Big) &=&
{1 \over 2 \lambda^2} \Gamma\Big(\Omega^-_{bb} \rightarrow \Xi^+_{bc} \
             (K^- \ K^-)_{L=even} \Big).
\end{eqnarray}
Two pairs of the above decay rates also hold regardless of $L$ value,
and odd values of $L$ add a decay rate to each: 
\begin{eqnarray}
\Gamma\Big( \Xi^-_{bb} \rightarrow \Omega^0_{bc} \ \pi^- K^0\Big) =
{1 \over \lambda^2} \ \Gamma\Big( \Omega^-_{bb} \rightarrow \Xi^0_{bc} 
 \ K^- \ \overline K^0 \Big) = 6 \ \Gamma\Big( \Xi^-_{bb} \rightarrow
 \Xi^0_{bc} \ (\pi^- \eta_8)_{L=odd}\Big),
\end{eqnarray}
\begin{eqnarray}
\Gamma\Big(\Omega^-_{bb} \rightarrow \Xi^+_{bc} \ \pi^- \ K^-
  \Big) = {1 \over \lambda^2} \ 
   \Gamma\Big(\Xi^-_{bb} \rightarrow \Xi^+_{bc} \ \pi^- \ K^-
  \Big) =
6 \ \Gamma\Big( \Xi^0_{bb} \rightarrow \Xi^+_{bc} \ (\pi^- \ \eta_8)_{L=odd}
  \Big).
\end{eqnarray}

Additional examples of relationships between Cabbibo-allowed and
Cabbibo suppressed decays which hold regardless of $L$ value are:
\begin{eqnarray}
\Gamma\Big(\Xi^0_{bb} \rightarrow \Xi^0_{bc} \ \pi^+ \ \pi^- \Big) =
{1 \over \lambda^2} \ 
  \Gamma\Big(\Xi^0_{bb} \rightarrow \Omega^0_{bc} \ K^+ \ K^- \Big) ,
\nonumber \\
\Gamma\Big(\Omega^-_{bb} \rightarrow \Omega^0_{bc} \ K^0 \ K^- \Big) =
{1 \over \lambda^2} \ 
  \Gamma\Big(\Xi^-_{bb} \rightarrow \Xi^0_{bc} \ \pi^- \ 
         \overline K^0 \Big) ,
\nonumber \\
\Gamma\Big(\Xi^0_{bb} \rightarrow \Omega^0_{bc} \ \pi^- K^+ \Big) =
{1 \over \lambda^2} \ 
  \Gamma\Big(\Xi^0_{bb} \rightarrow \Xi^0_{bc} \ \pi^+ K^- \Big) .
\end{eqnarray}

Note, however, that these are unbroken SU(3) relationships. We have
not treated explicit violations which may result, for instance, from
final state interactions in the form of mixing with nearby
resonances, which is a known issue for the two meson final states \cite{sw1}.

Once again, analogous relationships hold for $3_{bc} \rightarrow 3_{cc} + M +
M^\prime$.  

%********************************************************
\subsection{$3_{bb} \rightarrow 6_b + D$ 
(Final states with a b=--1 6 baryon plus a D meson)}
%********************************************************

The $3_{bb}$ may also decay to a member of the $D$-meson antitriplet,
Eqn.~\ref{Dmes}, and a baryon from the 6 representation of
SU(3), whose particles are identified in Eqn.~\ref{6b}.
There exist only two ways of constructing a singlet using these
multiplets
and the operator in Eqn.~\ref{oct}, yielding only two reduced 
matrix elements in which ten amplitudes can be written when SU(3) symmetry
is exact.  To include linear breaking terms
we also use the operators in Eqn.~\ref{octss},
which increases the number of reduced matrix elements by four.
\begin{eqnarray}
{\cal H}_{LB}(3_{bb} \rightarrow 6_b + D) 
&=& A \ [\overline 3_{bb}]_i \ [6_b]^{ij} \ D_k \ H_{(1)}(8)^k_j
    + B \ [\overline 3_{bb}]_i \ [6_b]^{jk} \ D_k \ H_{(1)}(8)^i_j  \nonumber \\
&+& A_{LB} \ [\overline 3_{bb}]_i \ [6_b]^{ij} \ D_k \ H_{(2)}(8)^k_j 
    + B_{LB} \ [\overline 3_{bb}]_i \ [6_b]^{jk} \ D_k \ H_{(2)}(8)^i_j  \nonumber \\
&+& C_{LB} \ 
  [\overline 3_{bb}]_i \ [6_b]^{kl} \ D_j \ H(\overline{10})^{ij}_{kl} 
    + D_{LB} \ [\overline 3_{bb}]_i \ [6_b]^{kl} \ D_j \ H(27)^{ij}_{kl}.
\end{eqnarray}
Note that there is no $H(10)$ contribution to this decay because
of the symmetry of the $[6_b]$.
The amplitudes are shown in Table~\ref{6bD}.  Example
relationships under SU(3) are:
\begin{eqnarray}
\Gamma(\Xi^0_{bb} \rightarrow  \Sigma^-_b \ D^+) = 2 \
\Gamma(\Xi^0_{bb} \rightarrow  \Xi^-_{b2} \ D_s^+) = {1 \over \lambda^2} \
\Gamma(\Xi^0_{bb} \rightarrow  \Omega^-_b \ D_s^+).
\end{eqnarray}

\begin{table}[!thb]
\begin{center}
\mbox{\epsffile{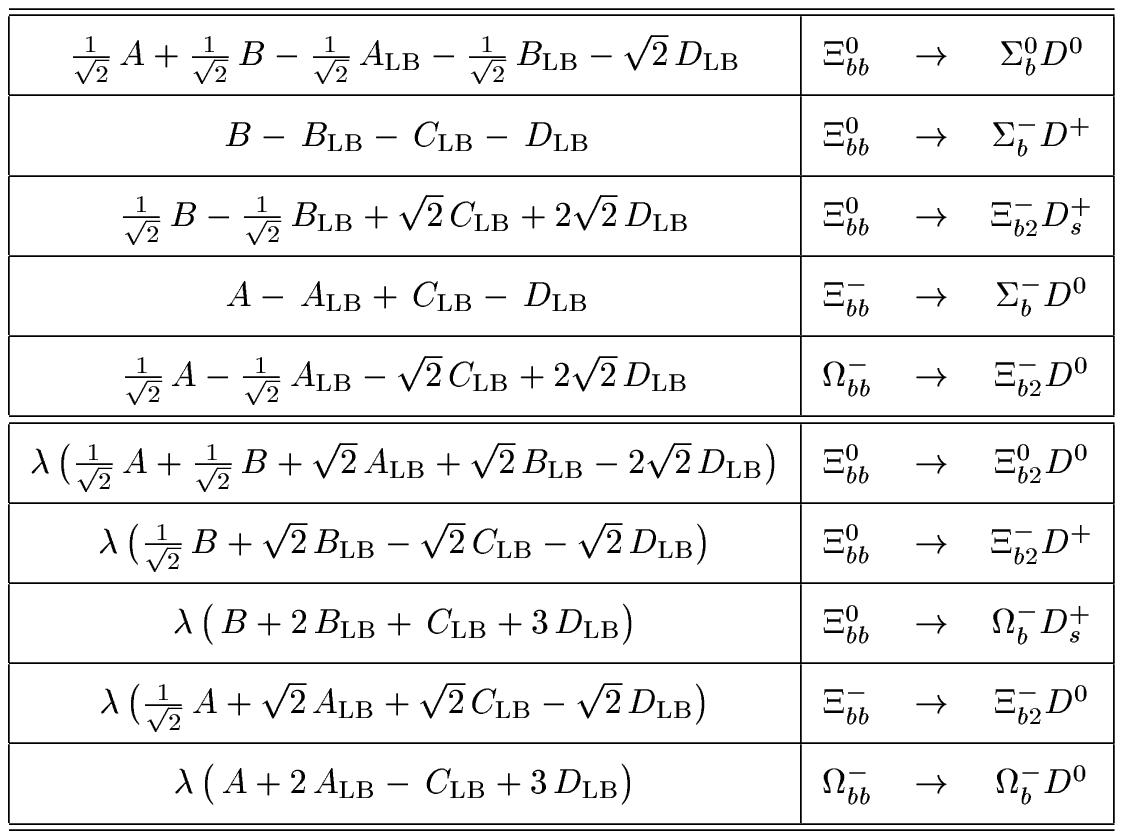}}
\caption{Matrix elements for the decay $3_{bb} \rightarrow 6_b + D$ \label{6bD}}
\end{center}
\end{table}

Including only linear breaking of SU(3) simplifies the ten processes, 
which are given in
terms of the ten reduced matrix elements required by arbitrary SU(3)
breaking in Appendix~\ref{3bb6bD}.  In that case, the 
reduced matrix elements behave as follows:
\begin{eqnarray}
\langle 15 || 27_{I=2} || 3 \rangle &=& 0, \nonumber\\
\langle 15 || 27_{I=\frac{3}{2}} || 3 \rangle &=& 0, \nonumber\\
\langle 15 || 10_{I=\frac{1}{2}} || 3 \rangle &=& 
   \lambda \ \langle 15 || 10_{I=1} || 3 \rangle , \nonumber\\
\langle 15 || 27_{I=\frac{1}{2}} || 3 \rangle &=& 
   \lambda \ \sqrt{3 \over 2} \ \langle 15 || 27_{I=1} || 3 \rangle.
\end{eqnarray}
Note that the $I=2$ and $I=\frac{3}{2}$  reduced matrix elements are again zero
and that the same pattern of CKM parameters relating
half-integer isospin to integer
isospin reduced matrix elements occurs in Eqn.~\ref{rel1}.
The analogous relationships for $3_{bc}$ decays are to the $6_c$ and
$D$.  The members of the $6_c$ are $\Omega_c^0$, $\Xi^+_{c2}$, $\Xi^0_{c2}$,
$\Sigma^{++}_c$, $\Sigma^+_c$, and $\Sigma_c^0$.

%********************************************************
\subsection{$3_{bb} \rightarrow \overline 3_b \ + \ D$ 
(Final states with an antitriplet baryon plus a D meson)}
%********************************************************

The last decay we will consider which takes place primarily through the 
operators given in Eqn.~\ref{oct} (or its extension in
Eqn.~\ref{octss}) is to a D meson plus
one  of the members of the antitriplet from Eqn.~\ref{3b}.

Including the linear SU(3) breaking terms the expression becomes:
\begin{eqnarray}
{\cal H}_{LB}(3_{bb} \rightarrow \overline 3_b + D) 
&=& A \ [\overline 3_{bb}]_i \ [\overline 3_{b}]_l \ 
      D_k \ H_{(1)}(8)^k_j  \ \epsilon^{lij}
    + B \ [\overline 3_{bb}]_i \ [\overline 3_{b}]_l \ 
      D_k \ H_{(1)}(8)^i_j  \ \epsilon^{ljk} \nonumber \\
&+& A_{LB} \ [\overline 3_{bb}]_i \ [\overline 3_{b}]_l \ 
      D_k \ H_{(2)}(8)^k_j  \ \epsilon^{lij}
    + B_{LB} \ [\overline 3_{bb}]_i \ [\overline 3_{b}]_l \ 
      D_k \ H_{(2)}(8)^i_j  \ \epsilon^{ljk} \nonumber \\
&+& C_{LB} \ [\overline 3_{bb}]_i \  [\overline 3_{b}]_m \ 
      D_j \ H(10)^{ij}_{kl}  \ \epsilon^{mkl}.
\end{eqnarray}

This decay does not admit of a term involving either the $\overline{10}$
or the $27$ because of the antisymmetry of the 
$[\overline 3_b]_i \epsilon^{ijk}$ structure.
The decay amplitudes are given in Table~\ref{3bD}.  
When SU(3) holds the amplitudes are related as follows:
\begin{eqnarray}
\Gamma \left( \Xi_{bb}^0 \rightarrow \Xi_{b1}^-\ D^+_s \right) &=& 
{1 \over \lambda^2}  \ 
\Gamma \left( \Xi_{bb}^0 \rightarrow \Xi_{b1}^-\ D^+ \right),
\nonumber \\
\Gamma \left( \Xi_{bb}^0 \rightarrow \Lambda_b^0 D^0 \right) &=&
{1 \over \lambda^2}  \ 
\Gamma \left( \Xi_{bb}^0 \rightarrow \Xi_{b1}^0\ D^0 \right),
\nonumber \\
\Gamma \left( \Omega_{bb}^- \rightarrow \Xi_{b1}^-\ D^0 \right) &=& 
{1 \over \lambda^2} \ 
\Gamma \left( \Xi_{bb}^- \rightarrow \Xi_{b1}^-\ D^0 \right).
\end{eqnarray}

\begin{table}[!thb]
\begin{center}
\mbox{\epsffile{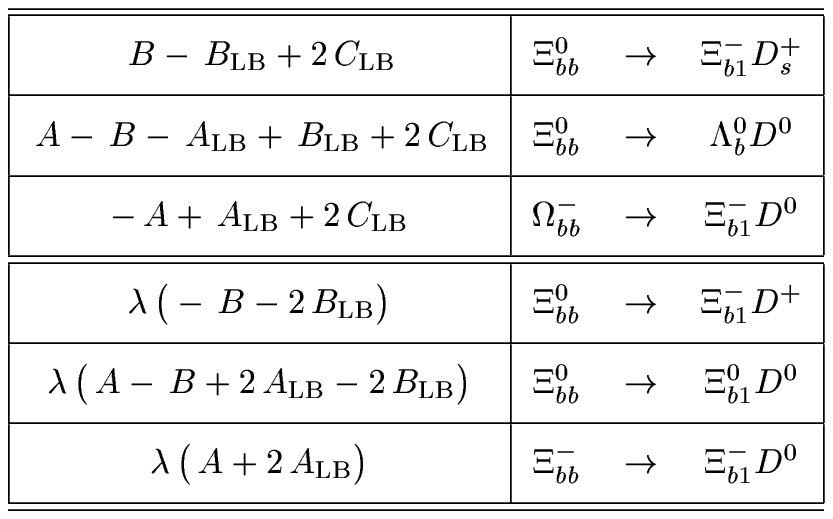}}
\caption{Matrix elements for the decay $3_{bb} \rightarrow \overline 3_b + D$ \label{3bD}}
\end{center}
\end{table}

Arbitrary SU(3) breaking yields six reduced matrix elements for the
six amplitudes of this type, shown in Appendix~\ref{3bbbar3bD}.  
When only linear breaking is
included, we find:
\begin{equation}
\langle \bar6 || \overline{10}_{I=\frac{3}{2}} || 3 \rangle = 0,
\end{equation}
because the Cabbibo suppressed operator $(b \overline c)(u \overline
s)(s \overline s)$ has no nonzero 10 components.  (Note that an
operator employed using the tensor methods described in the body of
this paper is expressed in terms of its barred components in the
arbitarily broken SU(3) reduced matrix elements used in Appendix
\ref{completedecomp}.  Therefore, the lack of 10 components for $(b
\overline c)(u \overline s)(s \overline s)$ corresponds to $\langle
\overline 6 || \overline{10}_{I=3/2} || 3\rangle = 0$.)  In this case,
we have six amplitudes expressed in terms of five reduced matrix elements.
Analogous relationships for the $3_{bc}$ decays are found from the
the table by making the appropriate $b \rightarrow c$ and charge replacement.
The final state particles are $\overline 3_c$ and $D$, where the
members of the $3_c$ are $\Lambda_c^+$, $\Xi^+_{c1}$, and $\Xi^0_c$.

%********************************************************
\subsection{$3_{bb} \rightarrow 3_{bc} \ + \ \overline D$
(Final states with a b=--1, c=1 triplet baryon plus a 
$\overline{\rm D}$ meson)}
%********************************************************

Now we treat the decays which utilize the operators given in Eqn.~\ref{atrip}
and Eqn.~\ref{light} and
their linear breaking extensions found in Eqns.~\ref{atripss}
and \ref{lightss3}-\ref{lightss15}.
The b=--2 SU(3) triplet particles in Eqn.~\ref{3bb} can decay 
through these operators to the triplet b=--1,
c=1 particles in Eqn.~\ref{3bc}, 
and an anti-$D$ meson.  There are six reduced matrix elements for the full
SU(3) case and two additional when linear breaking
is included.

\begin{table}[!thb]
\begin{center}
\mbox{\epsffile{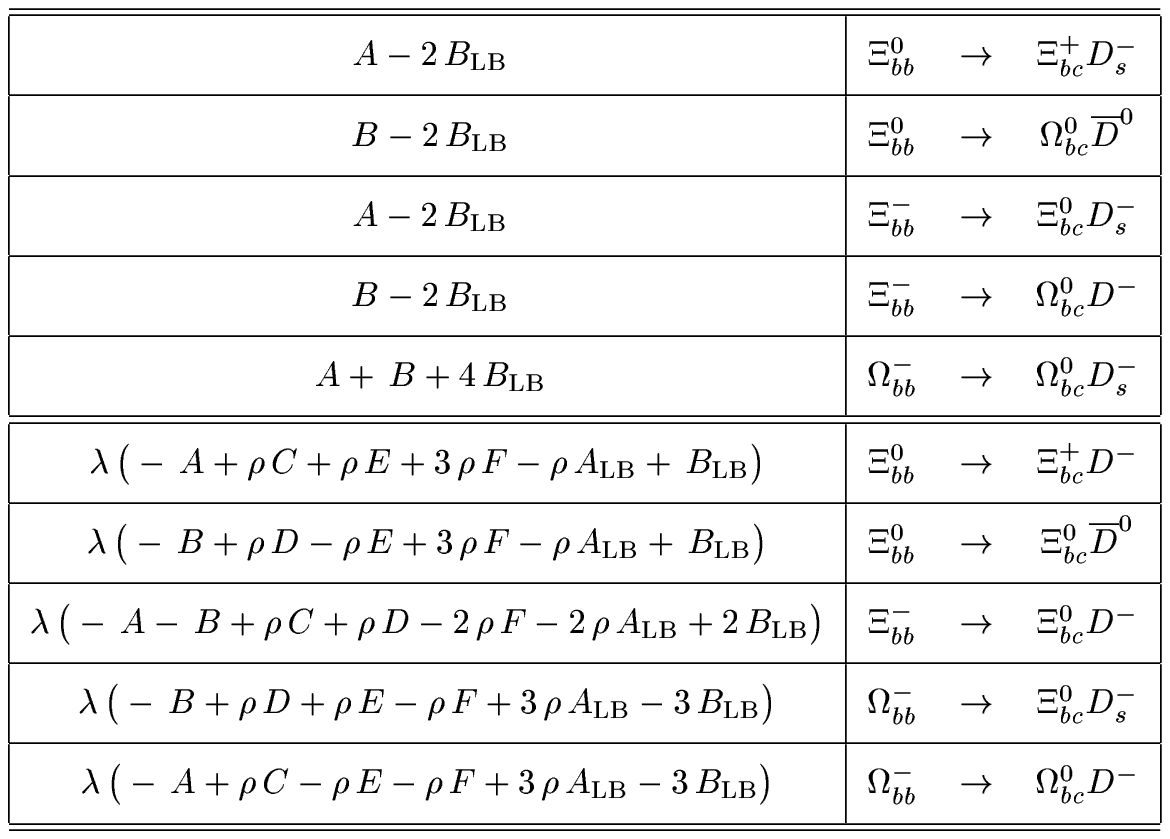}}
\caption{Matrix elements for the decay $3_{bb} \rightarrow 3_{bc} + \overline D$ \label{3bcD}}
\end{center}
\end{table}

\begin{eqnarray}\label{h3bcD}
{\cal H}_{LB}(3_{bb} \rightarrow 3_{bc} + \overline D) 
&=& A \ [\overline 3_{bb}]_i \ [ 3_{bc}]^i \ \overline D^j \ H_{(1)}(\overline 3)_j 
    + B \ [\overline 3_{bb}]_i \ [ 3_{bc}]^j \ \overline D^i \ H_{(1)}(\overline 3)_j 
  \nonumber \\
&+& C \ [\overline 3_{bb}]_i \ [ 3_{bc}]^i \ \overline D^j \ H''(\overline 3)_j 
    + D \ [\overline 3_{bb}]_i \ [ 3_{bc}]^j \ \overline D^i \ H''(\overline 3)_j 
  \nonumber \\
&+& E \ [\overline 3_{bb}]_i \ [ 3_{bc}]^j \ \overline D^k \ H^{\prime \prime}(6)^i_{jk} 
    + F \ [\overline 3_{bb}]_i \ [ 3_{bc}]^j \ \overline D^k \
   H^{\prime \prime}_{(1)}(\overline{15})^i_{jk}  
  \nonumber \\
&+& A_{LB} \ [\overline 3_{bb}]_i \ [ 3_{bc}]^j \ \overline D^k \
   H^{\prime \prime}_{(2)}(\overline{15})^i_{jk}
   + B_{LB} \ [\overline 3_{bb}]_i \ [ 3_{bc}]^j \ \overline D^k \
   H(\overline{15})^i_{jk}.
\end{eqnarray}
The results are shown in Table~\ref{3bcD}.  
With a linear breaking term included we have not only the $\overline{15}$
from the $(b \overline u) (u \overline d)$ operator,
$H_{(1)}^{\prime \prime}(\overline{15})$, but also 
$H_{(2)}^{\prime \prime}(\overline{15})$ as well as a different
$\overline{15}$ from $(b \overline c)(c \overline q)(s \overline s)$
of Eqn.~\ref{atripss}, $H(\overline{15})$.
Note that we do not need to include separate terms for $H(6)$ from
Eqn.~\ref{atripss} because 
the $H^{\prime\prime}(6)$ operator of Eqn.~\ref{light} 
already saturates this one (with the appropriate absorbtion of
$\rho$ factors).  For a similar reason, we do not need
$H_{(2)}(\overline 3)$, $H_{(3)}^{\prime \prime}(\overline{15})$, 
or $H_{(4)}^{\prime \prime}(\overline{15})$; the most general terms
are already contained in the operators we have included. 

Table~\ref{3bcD} shows that the following relationships hold not only
under unbroken SU(3) but also in the linear breaking case:
\begin{eqnarray}\label{bcbarDrate}
\Gamma \left(\Xi^0_{bb} \rightarrow \Xi^+_{bc} \ D^-_s \right) =
\Gamma \left(\Xi^-_{bb} \rightarrow \Xi^0_{bc} \ D^-_s \right)  
, \nonumber\\
\Gamma \left(\Xi^0_{bb} \rightarrow \Omega^0_{bc} \ \overline D^0 \right) =
\Gamma \left(\Xi^-_{bb} \rightarrow \Omega^0_{bc} \ D^- \right).
\end{eqnarray}

Comparing these results to Appendix~\ref{3bb3bcbarD} where arbitrarily
broken SU(3) gives the ten allowed processes in terms of the full
10 reduced matrix elements, the
reduced matrix elements behave as follows:
We have no isospin 1 elements because $(b \overline c)
(c \overline s)$ of Eqn~\ref{atrip} contains only $I=0$, 
and linear breaking with
$(s \overline s)$ does not modify this. The $I=1$ terms correspond
to the highly Cabbibo suppressed operator
neglected in our treatment. Therefore
\begin{eqnarray}\label{match3bcD}
\langle \overline 3 || \overline 6_{I=1} ||3 \rangle &=& 0 ,\nonumber\\
\langle 6 || \overline {15}_{I=1} ||3 \rangle &=& 0 ,
\end{eqnarray}
and we see how the ten unknown reduced matrix elements for arbitarily
broken SU(3) are reduced to the 8 in Eqn.~\ref{h3bcD}.
As in previous sections, the results for decays of the type
$3_{bc} \rightarrow 3_{cc} + \overline D$ are easily obtained from
Table~\ref{3bcD}.

%********************************************************
\subsection{$J/\Psi$ final states}
%********************************************************

The most promising decay modes for detecting the presence and decay of
doubly heavy baryons involves those with a $J/\Psi$ particle in the
final state since the subsequent $J/\Psi \rightarrow \mu^+ \mu^-$ decay
provides a clean experimental signature.

The $J/\Psi$ transforms as a singlet under SU(3).  It can occur in a final
state along with the antitriplet of Eqn.~\ref{3b} via the 
operators in Eqn.~\ref{atrip}
and in Eqn.~\ref{light} and their extensions in
Eqns.~\ref{atripss} and \ref{lightss3}-\ref{lightss15}.   The most general SU(3)
conserving form is unchanged by the addition of linear breaking terms:

\begin{eqnarray}
{\cal H}_{LB}(3_{bb} \rightarrow \overline 3_{b} + J/\Psi) 
&=& A \ [\overline 3_{bb}]_i \ [\overline 3_{b}]_l \ H_{(1)}(\overline 3)_j 
      \ \epsilon^{lij} 
 +  B \ [\overline 3_{bb}]_i \ [\overline 3_{b}]_l \
      H^{\prime\prime}(\overline 3)_j \ \epsilon^{lij} \nonumber\\
&+& C \ [\overline 3_{bb}]_i \ [\overline{3}_b]_j \ H^{\prime \prime}(6)^{ij},
\end{eqnarray}
because $H(6)^{ij}$ is already saturated in $H^{\prime \prime}(6)$ and  
$H_{(2)}(\overline 3)$ is contained in $H_{(1)}(\overline 3)$ and 
$H^{\prime\prime}(\overline 3)$.

The result is found in Table~\ref{jpsi3b} and 
\begin{eqnarray}
\Gamma \left( \Xi^0_{bb} \rightarrow \Xi^0_{b1} \ J/ \Psi \right) =
\Gamma \left( \Xi^-_{bb} \rightarrow \Xi^-_{b1} \ J/ \Psi \right)
\end{eqnarray} 
is a robust prediction, albeit a result of isospin.

\begin{table}[!thb]
\begin{center}
\mbox{\epsffile{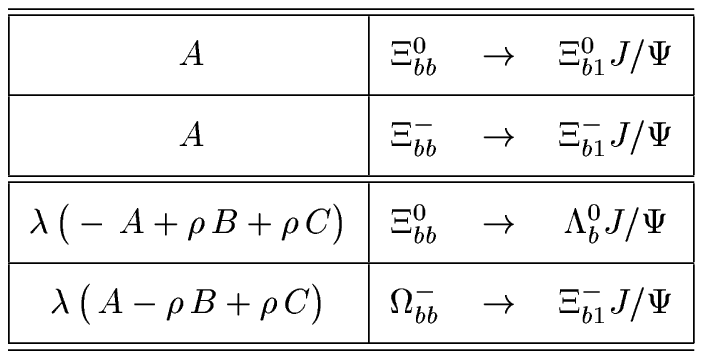}}
\caption{Matrix elements for the decay $3_{bb} \rightarrow \overline 3_{b} + J/\Psi$ \label{jpsi3b}}
\end{center}
\end{table}

When we compare this treatment with the result in Appendix~\ref{3bbbar3bJP}, 
which gives the decay amplitudes in terms of the
most general SU(3) breaking expressions, we find that:
\begin{equation}
\langle \bar 3 || \bar 6_{I=1} || 3\rangle = 0.
\end{equation}
This matrix element corresponds to a highly
Cabbibo suppressed operator, which we have neglected.  Now  we have
four amplitudes given in terms of three reduced matrix elements.

The $J/\Psi$ can also occur in a decay with a 6 from Eqn.~\ref{6b}
in the final state.  Three unknown reduced matrix elements 
come from the unbroken
SU(3) case, and two additional from the linear breaking term.
\begin{eqnarray}
{\cal H}_{LB}(3_{bb} \rightarrow 6_b + J/\Psi) &=&
A \ [\overline 3_{bb}]_i \ [6_b]^{ij} \ H_{(1)}(\overline 3)_j  +
B \ [\overline 3_{bb}]_i \ [6_b]^{ij} \ H''(\overline 3)_j  \nonumber\\
&+&C \ [\overline 3_{bb}]_i \ [6_b]^{jk} \ H_{(1)}^{\prime
\prime}(\overline{15})^i_{jk} 
\nonumber \\ &+&
A_{LB} \ [\overline 3_{bb}]_i \ [6_b]^{jk} \ H_{(2)}^{\prime
\prime}(\overline{15})^i_{jk} +
B_{LB} \ [\overline 3_{bb}]_i \ [6_b]^{jk} \ H(\overline{15})^i_{jk} \ ,
\end{eqnarray}
with results shown in Table~\ref{jpsi6b}. There is no term containing 
$H(6)^i_{jk}$, because of the antisymmetry in $j$ and $k$. And, as
discussed before, the above expression already saturates terms 
involving $H_{(2)}(\overline 3)$, $H_{(3)}^{\prime\prime}(\overline{15})$
and $H_{(4)}^{\prime\prime}(\overline{15})$. 

\begin{table}[!thb]
\begin{center}
\mbox{\epsffile{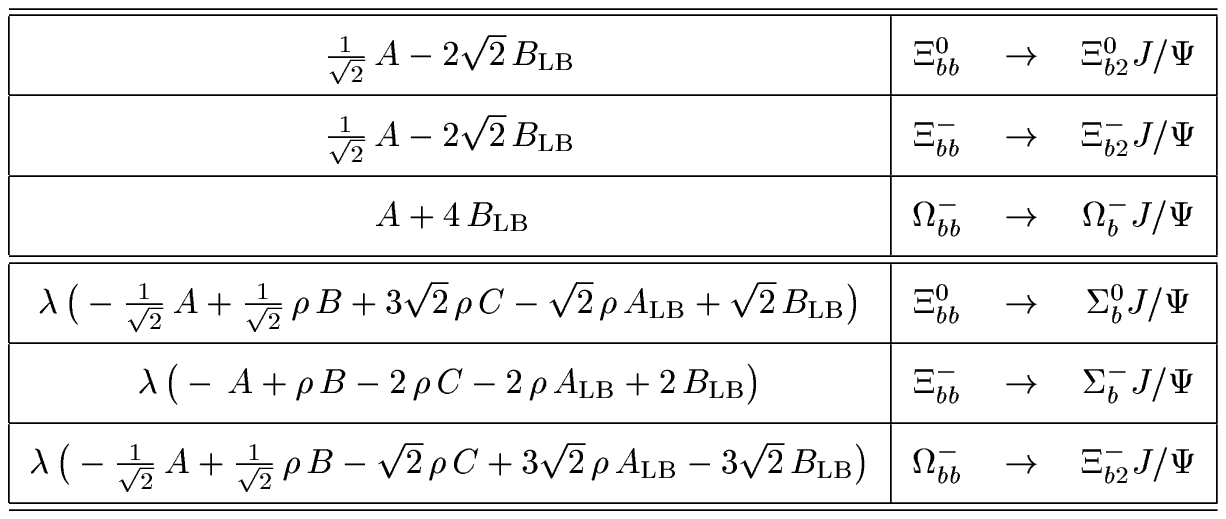}}
\caption{Matrix elements for the decay $3_{bb} \rightarrow 6_{b} +  J/\Psi$\label{jpsi6b}}
\end{center}
\end{table}

In the case of unbroken SU(3) we have
\begin{eqnarray}\label{6bjpsirate}
\Gamma \left( \Xi_{bb}^0 \rightarrow \Xi^0_{b2}\ J/\Psi \right) &=&
\Gamma \left( \Xi_{bb}^- \rightarrow \Xi^-_{b2}\ J/\Psi \right) =
{1 \over 2} \
\Gamma \left( \Omega_{bb}^- \rightarrow \Omega^-_{b}\ J/\Psi \right) 
, \nonumber\\
\Gamma \left( \Xi_{bb}^- \rightarrow \Sigma^-_{b}\ J/\Psi \right) &=&
2 \ \Gamma \left( \Omega_{bb}^- \rightarrow \Xi^-_{b2}\ J/\Psi \right).
\end{eqnarray}
Only the first equality survives when linear breaking terms are included.

Comparing the arbitrarily broken SU(3) reduced matrix
elements as given in Appendix~\ref{3bb6bJP} to the linear broken results
we find 
\begin{equation}
\langle 6 || 15_{I=1} || 3 \rangle = 0,
\end{equation}
which also follows from the fact that the integer isospin 
$\overline{15}$ comes from the
operator $(b\bar{c})(c\bar{s})$  in Eqn. \ref{atrip}, 
which is purely isospin 0.

Analogous results are obtained, if the $b$ quark decays first,
for the decays $3_{bc} \rightarrow \overline 3_c (6_c) J/\Psi$.

%********************************************************
\subsection{$3_{bb} \rightarrow \overline 3_b + M$ 
(Final states with a b=--1 antitriplet and an octet meson)}
%********************************************************

The operators shown in Eqn.~\ref{atrip} and Eqn.~\ref{light} 
also induce decays to a final antitriplet
of b=--1 baryons and an octet meson.  In the SU(3) symmetric case
we have eight reduced matrix elements. 
Including linear SU(3) breaking terms gives six additional 
from the available $\overline{15}$'s and $24$'s.
And so we have
\begin{eqnarray}
{\cal H}_{LB}(3_{bb} \rightarrow \overline 3_b + M)&=& \nonumber\\ 
&&\hspace{-2cm}+ A \ [\overline 3_{bb}]_i \ [\overline 3_b]_m \ M^k_j \ H_{(1)}(\bar 3)_k 
      \ \epsilon^{ijm}
    + B \ [\overline 3_{bb}]_i \ [\overline 3_b]_m \ M^i_j \ H_{(1)}(\bar 3)_k 
      \ \epsilon^{kjm}\nonumber \\
&&\hspace{-2cm}+ C \ [\overline 3_{bb}]_i \ [\overline 3_b]_m \ M^k_j \ H''(\bar 3)_k 
      \ \epsilon^{ijm}
    + D \ [\overline 3_{bb}]_i \ [\overline 3_b]_m \ M^i_j \ H''(\bar 3)_k 
      \ \epsilon^{kjm}\nonumber \\
&&\hspace{-2cm}+ E \ [\overline 3_{bb}]_i \  [\overline 3_b]_j \ M^i_l \ H^{\prime \prime}(6)^{jl} 
    + F \ [\overline 3_{bb}]_i \  [\overline 3_b]_j \ M^j_l \ H^{\prime \prime}(6)^{il} 
    \nonumber \\
&&\hspace{-2cm}+ G \ [\overline 3_{bb}]_i \  [\overline 3_b]_m \ M^k_l \ H_{(1)}^{\prime \prime}(\overline{15})^l_{jk} 
    \ \epsilon^{ijm} 
    + H \ [\overline 3_{bb}]_i \  [\overline 3_b]_m \ M^l_j \ 
    H_{(1)}^{\prime \prime}(\overline{15})^i_{kl} 
    \ \epsilon^{jkm}\nonumber \\
&&\hspace{-2cm}+ A_{LB} \ [\overline 3_{bb}]_i \ [\overline 3_b]_m \ M^k_l \ 
    H_{(2)}^{\prime \prime}(\overline{15})^l_{jk} \ \epsilon^{ijm}
    + B_{LB} \ [\overline 3_{bb}]_i \ [\overline 3_b]_m  \ M^l_j \ 
    H_{(2)}^{\prime \prime}(\overline{15})^i_{kl} \ \epsilon^{jkm}\nonumber \\
&&\hspace{-2cm}+ C_{LB} \ 
    [\overline 3_{bb}]_i \ [\overline 3_b]_m  \ M^k_l \ H(\overline{15})^l_{jk} 
    \ \epsilon^{ijm}
    + D_{LB} \ 
    [\overline 3_{bb}]_i \ [\overline 3_b]_m  \ M^l_j \ H(\overline{15})^i_{kl} 
    \ \epsilon^{jkm}\nonumber \\
&&\hspace{-2cm}+ E_{LB} \  
    [\overline 3_{bb}]_i \ [\overline 3_b]_j \ M^l_k \ 
    H_{(1)}^{\prime \prime}(24)^{ijk}_l 
    + F_{LB} \  
    [\overline 3_{bb}]_i \ [\overline 3_b]_j \ M^l_k \ 
    H_{(2)}^{\prime \prime}(24)^{ijk}_l.
\end{eqnarray}
\begin{table}[!thb]
\begin{center}
\mbox{\epsffile{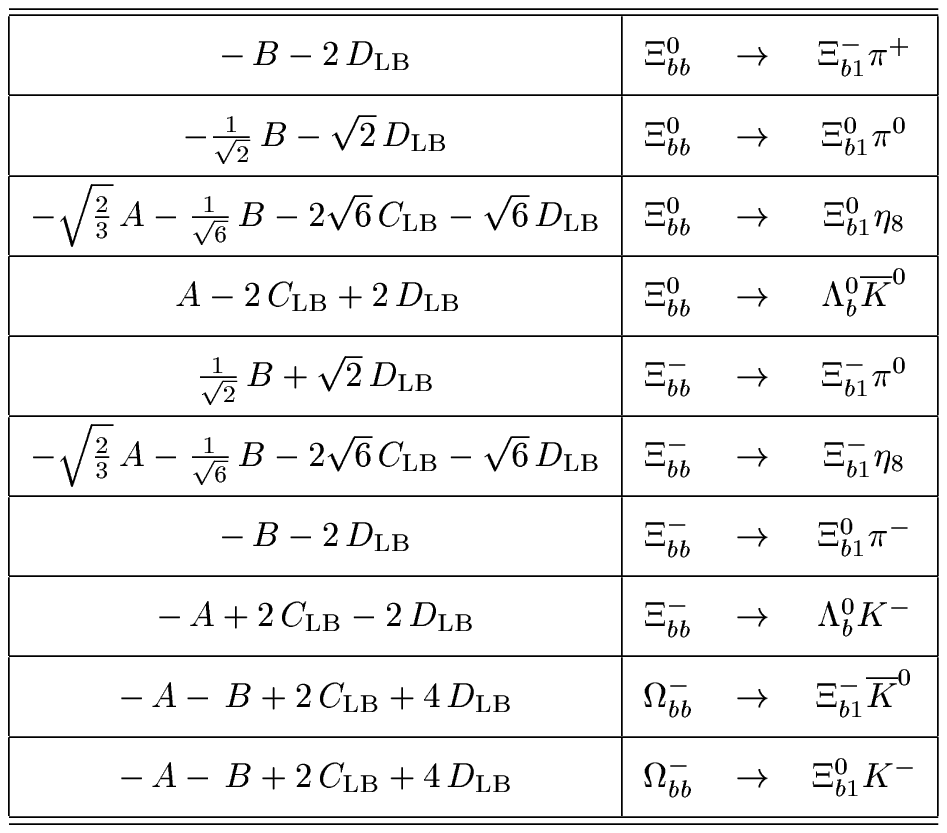}}
\caption{Matrix elements for the decay $3_{bb} \rightarrow \overline 3_b + M$; Cabbibo allowed decays
   \label{3bMa}}
\end{center}
\end{table}
From Tables~\ref{3bMa} and \ref{3bMb} we see that 
the following relationships between decay amplitudes hold:
\begin{eqnarray}
\Gamma \left( \Xi^0_{bb} \rightarrow \Xi^-_{b1} \ \pi^+ \right) &=&
2 \ \Gamma \left( \Xi^0_{bb} \rightarrow \Xi^0_{b1} \ \pi^0 \right) =
2 \ \Gamma \left( \Xi^-_{bb} \rightarrow \Xi^-_{b1} \ \pi^0 \right) =
\Gamma \left( \Xi^-_{bb} \rightarrow \Xi^0_{b1} \ \pi^- \right) 
, \nonumber\\
\Gamma \left( \Xi^0_{bb} \rightarrow \Lambda^0_{b} \ \overline K^0
\right) &=&
\Gamma \left( \Xi^-_{bb} \rightarrow \Lambda^0_{b} \ K^- \right) 
, \nonumber\\
\Gamma \left( \Xi^0_{bb} \rightarrow \Xi^0_{b1} \ \eta_8 \right) &=&
\Gamma \left( \Xi^-_{bb} \rightarrow \Xi^-_{b1} \ \eta_8 \right) 
, \nonumber\\
\Gamma \left( \Omega^-_{bb} \rightarrow \Xi^-_{b1} \ \overline K^0
\right) &=&
\Gamma \left( \Omega^-_{bb} \rightarrow \Xi^0_{b1} \ K^- \right).
\end{eqnarray}
Comparing this to the results given in Appendix~\ref{3bbbar3bM} we see that
restricting SU(3) breaking to a linear term gives us the following:
All six non-zero integer isospin reduced matrix elements are zero, for
the same reasons as given in previous sections.  Therefore we see how the 20 
reduced matrix elements are collapsed to 14
in the case of linear breaking.

As in previous sections, results for the decay $3_{bc} \rightarrow
\overline 3_c + M$ are obtained directly.

\begin{table}[!thb]
\begin{center}
\mbox{\epsffile{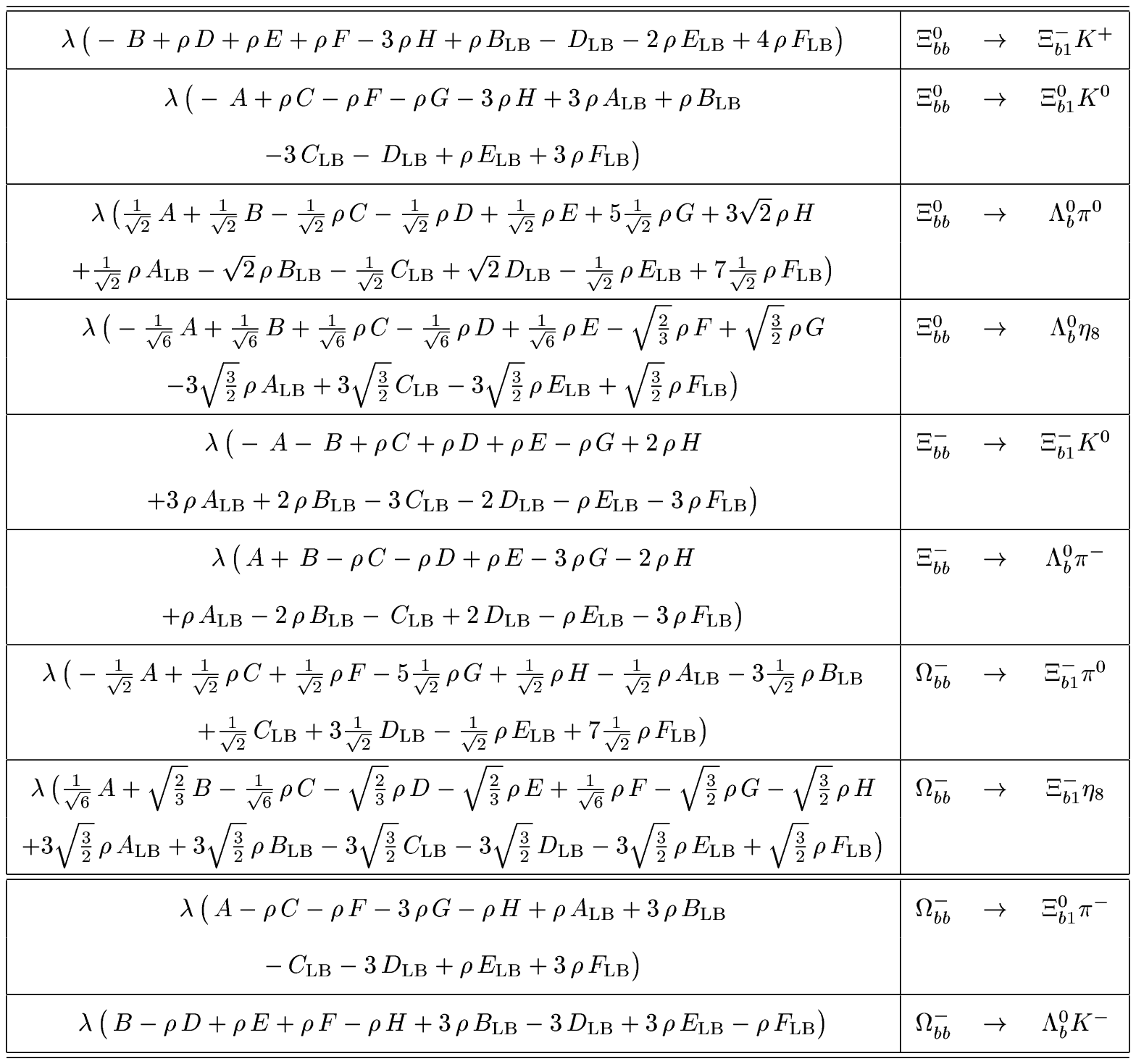}}
\caption{Matrix elements for the decay $3_{bb} \rightarrow \overline 3_b + M$; Cabbibo suppressed decays
   \label{3bMb}}
\end{center}
\end{table}

%********************************************************
\subsection{$3_{bb} \rightarrow 6_b + M$ 
(Final states with a b=--1 6 baryon plus an octet meson)}
%********************************************************

The operators shown in Eqn.~\ref{atrip} and Eqn.~\ref{light} 
also induce decays to a final $6$ 
of b=--1 baryons and an octet meson.  In the SU(3) symmetric case
we have nine reduced matrix elements.   
Including linear breaking terms gives us ten additional ones from
contraction with $H_{(2)}''(\overline{15})$, $H(\overline{15})$,  
$H''(\overline{15}^\prime)$, $H_{(1)}''(24)$, $H_{(2)}''(24)$ and
$H''(\overline{42})$.

\begin{eqnarray}
{\cal H}_{LB}(3_{bb} \rightarrow 6_b + M) &=& \nonumber\\
&&\hspace{-2.5cm}+ A \ [\overline 3_{bb}]_i \ [6_{b}]^{ij} \ M^k_j 
  \ H_{(1)}(\overline 3)_k 
   + B \ [\overline 3_{bb}]_i \ [6_{b}]^{jk} \ M^i_j 
  \ H_{(1)}(\overline 3)_k \nonumber \\
&&\hspace{-2.5cm}+ C \ [\overline 3_{bb}]_i \ [6_{b}]^{ij} \ M^k_j 
  \ H''(\overline 3)_k 
   + D \ [\overline 3_{bb}]_i \ [6_{b}]^{jk} \ M^i_j 
  \ H''(\overline 3)_k \nonumber \\
&&\hspace{-2.5cm}+ E \ [\overline 3_{bb}]_i \ [6_{b}]^{ij} \ M^k_l 
  \ H^{\prime \prime}(6)^l_{jk} 
   + F \ [\overline 3_{bb}]_i \ [6_{b}]^{jk} \ M^l_j 
  \ H^{\prime \prime}(6)^i_{kl} \nonumber \\
&&\hspace{-2.5cm}+ G \ [\overline 3_{bb}]_i \ [6_{b}]^{ij} \ M^k_l 
  \ H_{(1)}^{\prime \prime}(\overline{15})^l_{jk}
   + H \ [\overline 3_{bb}]_i \ [6_{b}]^{jk} \ M^l_j 
  \ H_{(1)}^{\prime \prime}(\overline{15})^i_{kl} \nonumber \\
&&\hspace{-2.5cm}+ I \ [\overline 3_{bb}]_i \ [6_{b}]^{jk} \ M^i_l 
  \ H_{(1)}^{\prime \prime}(\overline{15})^l_{jk} \nonumber \\
&&\hspace{-2.5cm}+ A_{LB} \ [\overline 3_{bb}]_i \ [6_{b}]^{ij} \ M^k_l 
  \ H_{(2)}^{\prime \prime}(\overline{15})^l_{jk}
  +B_{LB} \ [\overline 3_{bb}]_i \ [6_{b}]^{jk} \ M^l_j 
  \ H_{(2)}^{\prime \prime}(\overline{15})^i_{kl} \nonumber \\
&&\hspace{-2.5cm}+ C_{LB} \ [\overline 3_{bb}]_i \ [6_{b}]^{jk} \ M^i_l 
  \ H_{(2)}^{\prime \prime}(\overline{15})^l_{jk}
  + D_{LB} \ 
  [\overline 3_{bb}]_i \ [6_{b}]^{ij} \ M^k_l \ H(\overline{15})^l_{jk} \nonumber\\
&&\hspace{-2.5cm}+ E_{LB} \ 
  [\overline 3_{bb}]_i \ [6_{b}]^{jk} \ M^l_j \ H(\overline{15})^i_{kl} 
  + F_{LB} \ 
  [\overline 3_{bb}]_i \ [6_{b}]^{jk} \ M^i_l \ H(\overline{15})^l_{jk} \nonumber \\
&&\hspace{-2.5cm}+ G_{LB} \ 
  [\overline 3_{bb}]_i \ [6_{b}]^{jk} \ M^l_m \ 
   H^{\prime \prime} (\overline{15}^\prime)_{njkl} \ \epsilon^{imn}
  + H_{LB} \ 
   [\overline 3_{bb}]_i \ [6_{b}]^{jk} \ M^l_m 
  \ H_{(1)}^{\prime \prime}(24)^{imn}_j \ \epsilon_{nkl}\nonumber \\
&&\hspace{-2.5cm}+ I_{LB} \ 
   [\overline 3_{bb}]_i \ [6_{b}]^{jk} \ M^l_m 
  \ H_{(2)}^{\prime \prime}(24)^{imn}_j \ \epsilon_{nkl}
  +J_{LB} \ 
 [\overline 3_{bb}]_i \ [6_{b}]^{jk} \ M^l_m 
  \ H^{\prime \prime}(\overline{42})^{im}_{jkl}. 
\end{eqnarray}

The results are shown in Appendix~\ref{3bb6bMtable}.  
The following relationships
hold for unbroken SU(3):

\begin{eqnarray}
\Gamma \left( \Xi^0_{bb} \rightarrow \Sigma^+_b \ K^- \right) &=&
2 \ \Gamma \left( \Xi^0_{bb} \rightarrow \Sigma^0_b \ \overline K^0
\right) =
2 \ \Gamma \left( \Xi^-_{bb} \rightarrow \Sigma^0_b \ K^- \right) =
\Gamma \left( \Xi^-_{bb} \rightarrow \Sigma^-_b \ \overline K^0
\right), 
\nonumber\\
\Gamma \left( \Xi^0_{bb} \rightarrow \Xi^0_{b2 } \ \pi^0 \right) &=&
{1 \over 2} \ 
\Gamma \left( \Xi^0_{bb} \rightarrow \Xi^-_{b2} \ \pi^+ \right) =
{1 \over 4} \ 
\Gamma \left( \Xi^0_{bb} \rightarrow \Omega^-_b \ K^+ \right) =
{1 \over 2} \ 
\Gamma \left( \Xi^-_{bb} \rightarrow \Xi^0_{b2} \ \pi^- \right)
\nonumber\\
&=& \Gamma \left( \Xi^-_{bb} \rightarrow \Xi^-_{b2} \ \pi^0 \right) =
{1 \over 4} \ 
\Gamma \left( \Xi^-_{bb} \rightarrow \Omega^-_b \ K^0 \right),
\nonumber\\
\Gamma \left( \Xi^0_{bb} \rightarrow \Sigma^-_b \ \pi^+ \right) &=&
2 \ \Gamma \left( \Xi^0_{bb} \rightarrow \Xi^-_{b2} \ K^+ \right),
\nonumber\\
\Gamma \left( \Xi^0_{bb} \rightarrow \Xi^0_{b2} \ \eta_8 \right) &=&
\Gamma \left( \Xi^-_{bb} \rightarrow \Xi^-_{b2} \ \eta_8 \right),
\nonumber\\
\Gamma \left( \Omega^-_{bb} \rightarrow \Xi^0_{b2} \ K^- \right) &=&
\Gamma \left( \Omega^-_{bb} \rightarrow \Xi^-_{b2} \ \overline K^0
\right) =
{3 \over 4} \ \Gamma \left( \Omega^-_{bb} \rightarrow \Omega^-_b \ 
\eta_8 \right).
\end{eqnarray}
For linear breaking, the first line of relationships survives; the 
second (wrapping to third) breaks up into pairs of related
processes;
the fourth no longer holds; the fifth is obeyed, and in the six
line only the first equality holds.

Comparing to Appendix~\ref{3bb6bM}, of the 32 reduced matrix elements in
that table only nineteen survive after the following restrictions
from simple linear breaking: As before, all nonzero integer isospin reduced
matrix elements are zero.  That eliminates 10 reduced matrix elements.
There is no isospin $\frac{5}{2}$ component from Eqn~\ref{atrip} or \ref{light},
so $\langle 24 || 42_{I=\frac{5}{2}} || 3 \rangle$ is also eliminated.
Eqn.~\ref{atripss} does not contain a 42, so 
$\langle 24 || 42_{I=0} || 3 \rangle$ is zero.  
Finally, we find
\begin{equation}
\frac{5}{2\sqrt{2}} \ \langle 24 || 42_{I=\frac{1}{2}} ||3 \rangle = 
  \langle 24 || 42_{I=\frac{3}{2}} || 3 \rangle.
\end{equation}

If the $b$ quark decays first, we obtain similar results for the processes
$3_{bc} \rightarrow 6_c + M$.

%********************************************************
\subsection{$3_{bb} \rightarrow B + b$
(Final states with a B meson and a member of the baryon
octet)}
%********************************************************

The last decay types we will consider from the operators in
Eqns.~\ref{atrip}, \ref{atripss}, \ref{light}, and \ref{lightss3}-\ref{lightss15}
are those with $B$ mesons in the final state.  Using the 
multiplets given in Eqn.~\ref{Bmes} and \ref{boctet} we have

\begin{eqnarray}
{\cal H}_{LB}(3_{bb} \rightarrow B + b) &=&\nonumber\\
&&\hspace{-2cm}+ A \ [\overline 3_{bb}]_i \ B_j \ b^k_l \ H_{(1)}(\overline 3)_k \ 
    \epsilon^{ijl}
   + B \ [\overline 3_{bb}]_i \ B_j \ b^i_l \ H_{(1)}(\overline 3)_k \ 
    \epsilon^{jlk} \nonumber \\
&&\hspace{-2cm}+ C \ [\overline 3_{bb}]_i \ B_j \ b^k_l \ H''(\overline 3)_k \ 
    \epsilon^{ijl}
   + D \ [\overline 3_{bb}]_i \ B_j \ b^i_l \ H''(\overline 3)_k \ 
    \epsilon^{jlk} \nonumber \\
&&\hspace{-2cm}+ E \ [\overline 3_{bb}]_i \ B_j \ b^i_l \ H^{\prime \prime}(6)^{jl}  
   + F \ [\overline 3_{bb}]_i \ B_j \ b^j_l \ H^{\prime \prime}(6)^{il}  
   \nonumber \\
&&\hspace{-2cm}+ G \ [\overline 3_{bb}]_i \ B_j \ b^m_l \ H_{(1)}^{\prime \prime}(\overline{15})^j_{km} 
  \ \epsilon^{kil}
  + H \ [\overline 3_{bb}]_i \ B_j \ b^m_l \ H_{(1)}^{\prime \prime}(\overline{15})^l_{km} 
  \ \epsilon^{kij} \nonumber \\
&&\hspace{-2cm}+ A_{LB} \ [\overline 3_{bb}]_i \ B_j \ b^m_l \ H_{(2)}^{\prime \prime}(\overline{15})^j_{km} 
  \ \epsilon^{kil}
  + B_{LB} \ [\overline 3_{bb}]_i \ B_j \ b^m_l \ H_{(2)}^{\prime \prime}(\overline{15})^l_{km} 
  \ \epsilon^{kij} \nonumber \\
&&\hspace{-2cm}+ C_{LB} \ [\overline 3_{bb}]_i \ B_j \ b^m_l \ H(\overline{15})^j_{km} 
  \ \epsilon^{kil} 
  + D_{LB} \ [\overline 3_{bb}]_i \ B_j \ b^m_l \ H(\overline{15})^l_{km} 
  \ \epsilon^{kij}  \nonumber \\ 
&&\hspace{-2cm}+ E_{LB} \ [\overline 3_{bb}]_i \ B_j \ b^k_l \ H_{(1)}^{\prime \prime}(24)^{ijl}_k 
  + F_{LB} \ [\overline 3_{bb}]_i \ B_j \ b^k_l \ H_{(2)}^{\prime \prime}(24)^{ijl}_k.
\end{eqnarray}

\begin{table}[!thb]
\begin{center}
\mbox{\epsffile{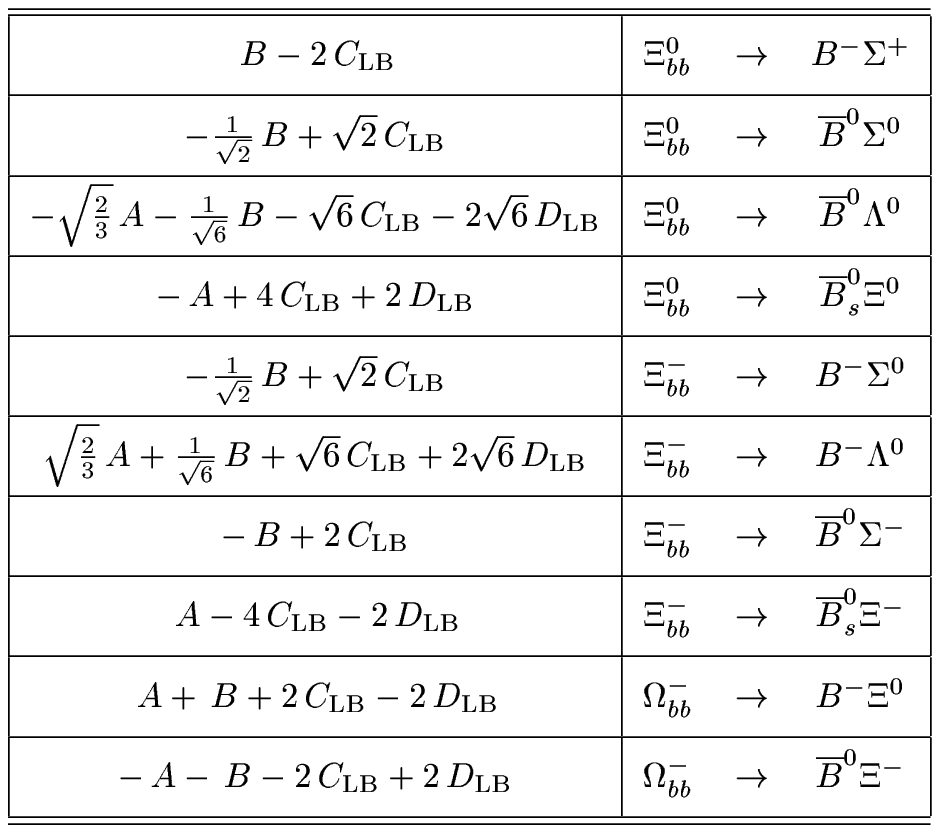}}
\caption{Matrix elements for the decay $3_{bb} \rightarrow B +  b$; Cabbibo allowed decays \label{Bba}}
\end{center}
\end{table}

Relationships are to be found in Tables~\ref{Bba} and \ref{Bbb}.  
Linearly broken SU(3)
yields:

\begin{eqnarray}
\Gamma \left( \Xi_{bb}^0 \rightarrow B^- \ \Sigma^+ \right) &=&
 2 \ 
  \Gamma \left( \Xi_{bb}^0 \rightarrow \overline B^0 \ \Sigma^0
 \right)  = 2 \ 
 \Gamma \left( \Xi_{bb}^- \rightarrow  B^- \ \Sigma^0 \right) =
\Gamma \left( \Xi_{bb}^- \rightarrow \overline B^0 \ \Sigma^- \right), 
\nonumber\\
\Gamma \left( \Xi_{bb}^0 \rightarrow \overline B^0 \ \Lambda^0 \right)
  &=&
\Gamma \left( \Xi_{bb}^- \rightarrow B^- \ \Lambda^0 \right),
\nonumber\\
\Gamma \left( \Xi_{bb}^0 \rightarrow \overline B^0_s \ \Xi^0 \right)
  &=& 
 \Gamma \left( \Xi_{bb}^- \rightarrow \overline B^0_s \ \Xi^-  \right),
\nonumber\\
\Gamma \left( \Omega_{bb}^- \rightarrow  B^- \ \Xi^0 \right) &=&
\Gamma \left( \Omega_{bb}^0 \rightarrow \overline B^0 \ \Xi^- \right).
\end{eqnarray}

In terms of the twenty arbitarily broken SU(3) reduced matrix elements from
Appendix~\ref{3bbBb} we see that the six nonzero integer isospin reduced
matrix elements are zero.

In this case, analogous relationships for $3_{bc}$ decay will be
found to final states $D$ plus octet baryon.

\begin{table}[!thb]
\begin{center}
\mbox{\epsffile{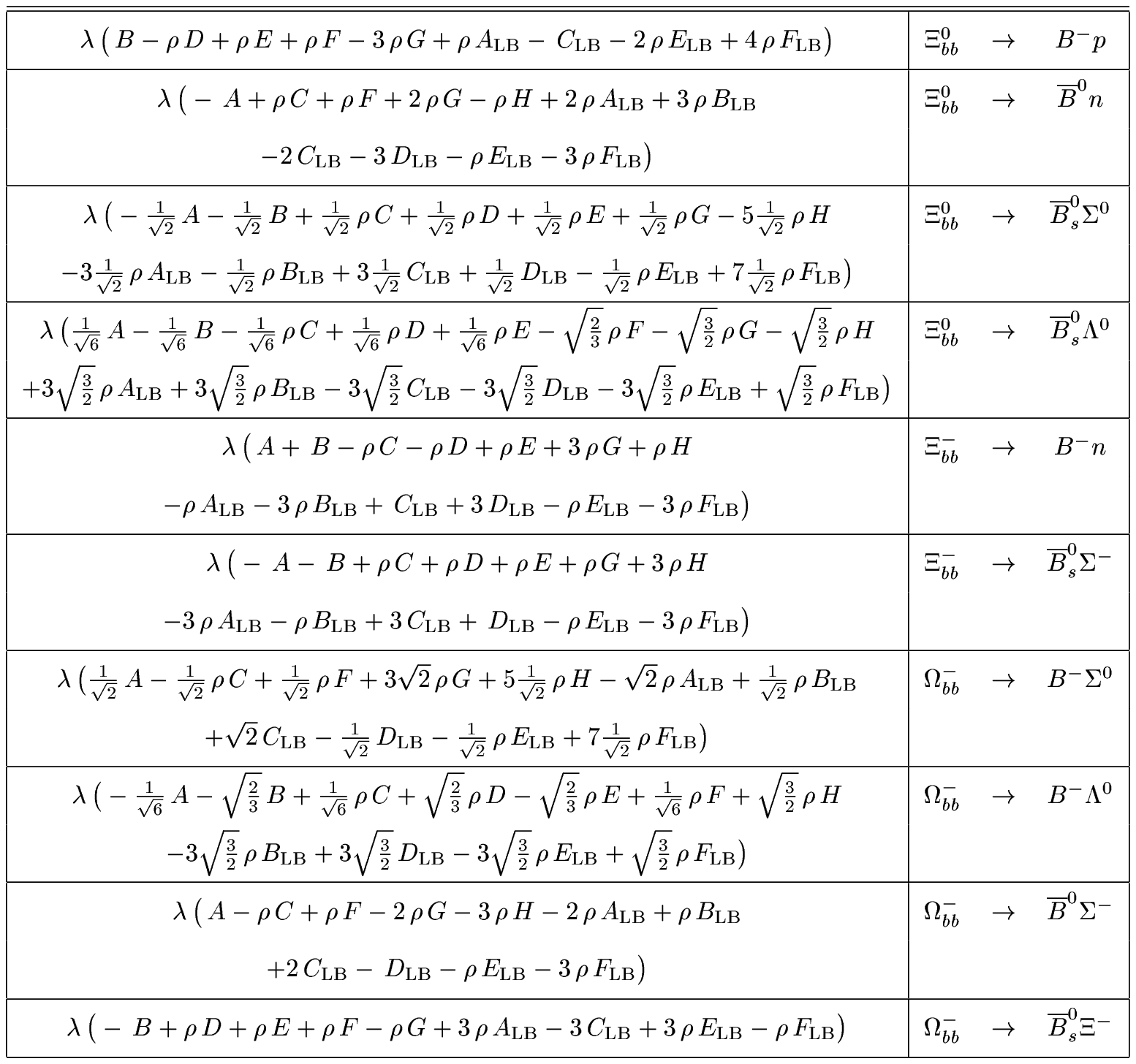}}
\caption{Matrix elements for the decay $3_{bb} \rightarrow B +  b$; Cabbibo suppressed decays \label{Bbb}}
\end{center}
\end{table}

\clearpage

%********************************************************
\subsection{$3_{bb} \rightarrow \overline 3_b + \overline D$
(Final states with an antitriplet $b$ baryon and a $\overline D$ 
in the final state)}
%********************************************************

The decay processes in these last two sections use the operators
in Eqns.~\ref{bucs} and \ref{bucslin}.
\begin{eqnarray}
{\cal H}_{LB}(3_{bb} \rightarrow \overline 3_b + \overline D) 
&=& A \ [\overline 3_{bb}]_i \ [\overline 3_b]_j \ \overline D^i \ H^\prime(3)^j  
    + B \ [\overline 3_{bb}]_i \ [\overline 3_b]_j \ \overline D^j \ H^\prime(3)^i  
   \nonumber \\
&+& C \ [\overline 3_{bb}]_i \ [\overline 3_{b}]_l \ \overline D^k \ H^\prime(\overline 6)_{jk}  
   \ \epsilon^{lij}
   + A_{LB} \ [\overline 3_{bb}]_i \ [\overline 3_b]_j \ \overline D^k \
  H^\prime(15)^{ij}_k.
\end{eqnarray}
The result is found in Table~\ref{3bDbar}.  Barring assumptions made
about phases, there are no obvious relationships even for unbroken
SU(3).

\begin{table}[!thb]
\begin{center}
\mbox{\epsffile{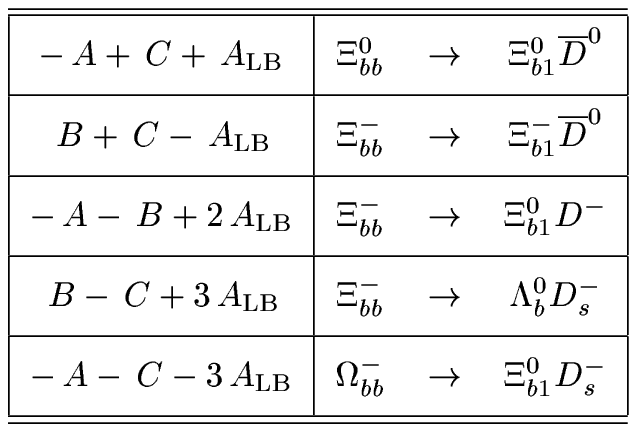}}
\caption{Matrix elements for the decay $3_{bb} \rightarrow \overline3_{b} +  \overline D$ \label{3bDbar}}
\end{center}
\end{table}

Comparing this to arbitarily broken SU(3), from Appendix~\ref{3bbbar3bbarD}, we
find that, if restricted to simply the linear breaking case,
only $\langle 8 || \overline{15}_{I=\frac{3}{2}} || 3 \rangle$ is
eliminated (since $(b \overline u) (c \overline s)$ is purely $I=\frac{1}{2}$).

A similar analysis applies to the process $3_{bc} \rightarrow \overline 3_c
+\overline D$.

%********************************************************
\subsection{$3_{bb} \rightarrow 6_b + \overline D$ (Final states
with a b=--1 6 baryon and a $\overline{D}$ in the final state)}
%********************************************************

\begin{eqnarray}
{\cal H}_{LB}(3_{bb} \rightarrow 6_b + \overline D) 
&=& A \ [\overline 3_{bb}]_i \  [6_b]^{ij} \ \overline D^k \ H^\prime(3)^l \ \epsilon_{ljk} 
    + B \ [\overline 3_{bb}]_i \ [6_b]^{ij} \ \overline D^k \ H^\prime(\overline 6)_{jk}  
  \nonumber \\
&+& C \ [\overline 3_{bb}]_i \ [6_b]^{jk} \ \overline D^i \ H^\prime(\overline 6)_{jk}  
  \nonumber \\
&+& A_{LB} \ 
  [\overline 3_{bb}]_i \ [6_b]^{jk} \ \overline D^l \ H^\prime(15)^{im}_k \ 
  \epsilon_{jlm}
  + B_{LB} \ [\overline 3_{bb}]_i \ [6_b]^{jk} \ \overline D^l \ 
  H^\prime(\overline{24})_{jkl}^i.\nonumber \\
\end{eqnarray}

\begin{table}[!thb]
\begin{center}
\mbox{\epsffile{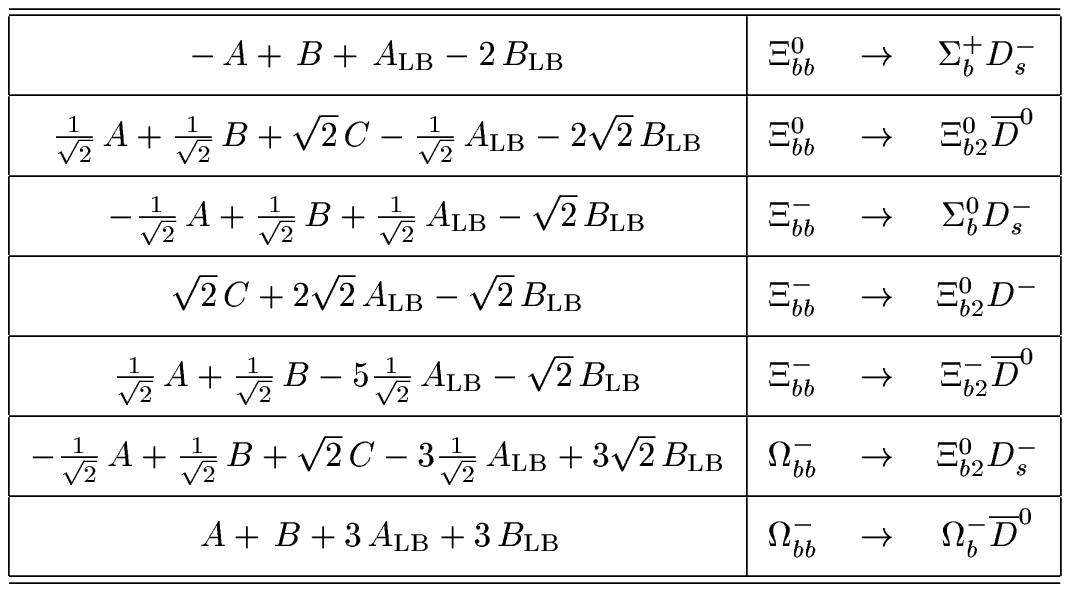}}
\caption{Matrix elements for the decay $3_{bb} \rightarrow 6_{b} +  \overline D$ \label{6bDbar}}
\end{center}
\end{table}

Results are given in Table~\ref{6bDbar}.  
According to Table~\ref{6bDbar} the following relationships hold in the
case of an exact $SU(3)$
\begin{eqnarray}
\Gamma \left(\XbbO\rightarrow \Sbp \ \Dsm\right) &=&
2 \ \Gamma \left(\Xbbm\rightarrow \SbO\ \Dsm\right),\nonumber\\
2 \ \Gamma \left(\Xbbm\rightarrow \XbZm \ \barDO\right) &=&
\Gamma \left(\Obbm\rightarrow \Obm \ \barDO\right),
\end{eqnarray}
where the first relation remains true in the linear broken case.

Comparing this to arbitarily broken SU(3), from Appendix~\ref{3bb6bbarD}, we
find that 
$\langle 8 ||\overline{15}_{I=\frac{3}{2}} ||3 \rangle = 0$ and
$\langle 10 ||24_{I=\frac{3}{2}} ||3 \rangle = 0$.

As before, the processes in the multiplet structure $3_{bc} \rightarrow
6_c + \overline D$ give analogous relationships.

\section{Summary}

In conclusion, we have examined various decay modes of the lowest
lying hadrons containing two $b$ quarks, with obvious extensions
to hadrons containing one $b$ and one $c$ quark.  We have provided the
predictions resulting from imposing exact SU(3) flavor symmetry,
and those arising from inclusion of a linear breaking piece in
the form of a strange quark mass.  The decomposition into
the group-theoretic basis, and the relationship between this
basis and the linear breaking case, should aid in further analyses
involving diagrammatic, factorization, or large $N_c$ treatments.
If the SU(3) relationships are measured to hold in the $b$ hadrons, this may
suggest that the violations found in the charm system are due to the influence
of nearby resonances unique to that system.  If the linear breaking
relationships hold, this may suggest that the breaking caused
by the strange quark mass can be treated perturbatively.  We hope
that these results will not only aid the experimentalists in finding
new doubly heavy hadrons, but will serve as a useful comparison to
those studying decay modes using other methods. Together these efforts
will help to illuminate the properties of heavy quark systems.

%********************************************************************************
\centerline{\bf Acknowledgments}
%********************************************************************************

\bigskip
{\samepage
RPS thanks C. Quigg and S. Ellis for requesting the calculation of
the SU(3) conserving predictions, and
for their invitations to the FNAL b Physics
Workshops.  RPS also thanks Z. Ligeti for convincing her that the
results should be present in the literature. This eventually led to the rest
of this work. RPS gratefully acknowledges the hospitality of the TU
and W. Weise's group. RPS acknowledges useful
discussions with T.J. Allen, T. Mehen, and C. Schat.  DAE thanks Duke
University for their hospitality while some of this work was done.
JU acknowledges useful discussions with S. Bosch and R.
Rosenfelder.
DAE acknowledges support from the Alfred P. Sloan Foundation and from
National Science Foundation Grant No. DMR-0094178.
%the National Science Foundation, grant no. DMR-9705410. 
This work was supported in part by the 
German `Bundesministerium f\"ur Bildung und Forschung' under contract 
05HT1WOA3 and by the `Deutsche Forschungsgemeinschaft' (DFG) under 
contract Bu.706/1-1 and Gr 1887/2-1 as well as from the US Department of Energy
under grant no.  DE-FG02-96ER40945.}

\clearpage

%********************************************************************************

\nopagebreak

%********************************************************************************

\begin{appendix}

%********************************************************************************
%****************************************************************************
\section{
\hbox{Matrix elements for the decay $3_{\lowercase{bb}}\rightarrow
3_{\lowercase{bc}}
      + M + M$,}
\hbox{even angular momentum $L$ ($SU(3)$ exact results).}}
\label{3bb3bcMMa}

\input{SU3exacta}

%****************************************************************************
\section{
\hbox{Matrix elements for the decay $3_{\lowercase{bb}}\rightarrow
3_{\lowercase{bc}}
      + M + M$,}
\hbox{odd angular momentum $L$ ($SU(3)$ exact results).}}
\label{3bb3bcMMb}
%****************************************************************************

%\input{pro3bb3bcMMb}
\input{SU3exactb}

%********************************************************************************
\section{Matrix elements for the decay $3_{\lowercase{bb}}\rightarrow 6_{\lowercase{b}} + M$}
\label{3bb6bMtable}
%********************************************************************************

\input{pro3bb6bM}

%********************************************************************************
\section{Tensor decomposition}
\label{tendec}
%********************************************************************************

In the following we present the tensor decompositions needed to
provide the elements given in Eqs. 
\ref{atrip}--\ref{ham4part3} of Sec. \ref{effham}.
While the normalization does not play any role in  Sec. \ref{effham}
we will keep it explicit here.
\begin{enumerate}
\item The operator $(b\bar{c})(c\bar{q})(s\bar{s})_8$  decomposes via
  $\bar{3}\otimes 8=
      \overline{15}\oplus 6\oplus \bar{3}$. Tensor methods yield
\cite{georgi:group}
      \begin{eqnarray}
      \bar{3}_i 8^j_k &=& 
      \underbrace{ \frac{1}{2}\left(\bar{3}_i 8^j_k+\bar{3}_k 8^j_i-
                   \frac{1}{4}\delta^j_i \bar{3}_l 8^l_k - 
                   \frac{1}{4}\delta^j_k \bar{3}_l 8^l_i\right)}_
                 {\overline{15}}
       +\epsilon_{ikl} \underbrace{\frac{1}{4}\left(
         \epsilon^{lmn} \bar{3}_m 8^j_n+\epsilon^{jmn} \bar{3}_m 8^l_n
         \right)}_{6}\nonumber\\
       &+&\underbrace{ \frac{1}{8} \left(3 \delta^j_i \bar{3}_l 8^l_k-
         \delta^j_k \bar{3}_l 8^l_i\right)}_{\bar 3}.      
       \label{b3x8}
       \end{eqnarray}
       Together with Eqs. \ref{atrip} and an explicit expression
       for $(s\bar s)_8$ we find (up to normalization)
       the numbers given in Eq. \ref{atripss}.  $(s\bar s)_8$ 
is found from
       \begin{equation}
       3^i \bar{3}_j = \underbrace{ \left(3^i \bar{3}_j-\frac{1}{3}
       \delta^i_j 3^l \bar{3}_l\right)}_{8}+\underbrace{ \frac{1}{3}
       \delta^i_j 3^l \bar{3}_l}_{1}
       \quad \Longrightarrow \quad
       (s\bar s)_8 = \left(\begin{array}{ccc}
       -\frac{1}{3} & 0 & 0\\0 & -\frac{1}{3} & 0\\ 0 & 0 & \frac{2}{3}
       \end{array}\right).
       \end{equation}
\item The operator $(b\bar{c})(u\bar{q})$ is contained in an 8-dimensional
      representation of $SU(3)$. Including linear breaking,
      $(s\bar{s})$, requires that $8\otimes 8'$ be decomposed.

      \begin{eqnarray}
      8^i_k {8'}^j_l &=& \underbrace{\frac{1}{4} \left( 8^i_k {8'}^j_l
      + {\cal P} - \frac{1}{5} \left\{ \delta^i_l 8^m_k {8'}^j_m +
      \delta^j_k 8^i_m {8'}^m_l + {\cal P} \right\} + \frac{1}{10}
      \left\{ \delta^i_l \delta^j_k + \delta^j_l \delta^i_k\right\}
      8^m_n {8'}^n_m \right)}_{27}\nonumber\\ &+&\epsilon_{klm}
      \underbrace{\frac{1}{12} \left( \epsilon^{mno} 8^i_n {8'}^j_o +
      {\cal P}\right)}_{10} + \epsilon^{ijm} \underbrace{\frac{1}{12}
      \left( \epsilon_{mno} 8^n_k {8'}^o_l + {\cal
      P}\right)}_{\overline{10}}\nonumber\\
      &&\!\!\!\!\!\!\!\underbrace{-\ \delta^i_k
      {8_{(sym)}}^j_l+\frac{3}{2} \delta^i_l {8_{(sym)}}^j_k +\frac{3}{2}
      \delta^j_k {8_{(sym)}}^i_l - \delta^j_l
      {8_{(sym)}}^i_k}_{8_{(a)}}\nonumber\\
      &&\!\!\!\!\!\!\!\underbrace{-\frac{5}{6} \delta^i_l
      {8_{(asym)}}^j_k +\frac{5}{6} \delta^j_k {8_{(asym)}}^i_l}_{8_{(b)}}
      +\underbrace{\frac{1}{8} \left(\delta^i_l \delta^j_k-
      \frac{1}{3} \delta^j_l \delta^i_k\right) 8^n_m {8'}^m_n}_{1}.
      \end{eqnarray}
      ${\cal P}$ stands for all possible permutations of the free
      upper and free lower indices. We have defined the following
      abbreviations %
      \begin{eqnarray}
      {8_{(sym)}}^j_k &=& \frac{1}{5}
                      \left(8^j_o 8'^o_k + 8^o_k 8'^j_o - 
                            \frac{2}{3} \delta^j_k 8^p_o 8'^o_p\right),
      \nonumber\\
      {8_{(asym)}}^j_k &=& \frac{1}{5}
                      \left(8^j_o 8'^o_k -8^o_k 8'^j_o\right).
      \end{eqnarray}
\item The operator $(b\bar u)(c\bar s)$  decomposes into
      a $3$ and a $\bar 6$. The numbers given in Eq. \ref{bucs} are 
found using
      \begin{equation}
      \bar{3}_i \bar{3}_j = \underbrace{\frac{1}{2} \left(\bar{3}_i \bar{3}_j+
      \bar{3}_j \bar{3}_i\right)}_{\bar 6} + \epsilon_{ijk} \underbrace{\frac{1}{2}
      \epsilon^{klm} \bar{3}_l \bar{3}_m}_{3}.
      \end{equation}
      Including a linear breaking term gives $(b\bar u)(c\bar s)(s\bar s)$, 
or $(\bar{6}\oplus 3)\otimes(8\oplus 1)$.
      We need the following tensor decompositions, where now ${\cal P}$
does not include permutations over the (symmetric) indices of $\bar{6}_{ij}$.
      \begin{eqnarray}\label{6x8}
      \bar{6}_{ij} 8^l_m &=& \underbrace{\frac{1}{3} \left(\bar{6}_{ij} 8^l_m 
      + {\cal P} -\frac{1}{5} \left\{ \delta^l_i \bar{6}_{nj} 8^n_m  + {\cal P}\right\}
      \right)}_{\overline{24}}\nonumber\\
      &+&\underbrace{\frac{1}{6} \epsilon_{inm} \left( 
      \bar{6}_{qj} 8^l_p \epsilon^{npq} + 
      \bar{6}_{qj} 8^n_p \epsilon^{lpq}
      - \frac{1}{4} \left\{
      \delta^n_j \bar{6}_{qo} 8^o_p \epsilon^{lpq} +
      \delta^l_j \bar{6}_{qo} 8^o_p \epsilon^{npq}
      \right\}\right)+
      i\leftrightarrow j}_{15}\nonumber\\
      &+&\underbrace{\frac{3}{20} \delta^l_i \left( \bar{6}_{mn} 8^n_j +
      \bar{6}_{jn} 8^n_m\right)+\frac{3}{20} \delta^l_j \left( \bar{6}_{mn} 8^n_i +
      \bar{6}_{in} 8^n_m\right)-\frac{1}{10} \delta^l_m \left( \bar{6}_{in} 8^n_j +
      \bar{6}_{jn} 8^n_i\right)}_{\bar{6}}\nonumber\\
      &+&\underbrace{\frac{1}{8} \left(\delta^l_i \epsilon_{mjo} +
      \delta^l_j \epsilon_{mio}\right) \bar{6}_{qr} 8^q_p
      \epsilon^{pro}}_{3}
      \end{eqnarray}
      and 
      \begin{eqnarray}
      3^i 8^j_k &=& 
      \underbrace{ \frac{1}{2}\left(3^i 8^j_k+3^j 8^i_k-
                   \frac{1}{4}\delta^i_k 3^l 8^j_l - 
                   \frac{1}{4}\delta^j_k 3^l 8^i_l\right)}_
                 {15}
       +\epsilon^{ijl} \underbrace{\frac{1}{4}\left(
         \epsilon_{lmn} 3^m 8^n_k+\epsilon_{kmn} 3^m 8^n_l
         \right)}_{\bar{6}}\nonumber\\
       &+&\underbrace{ \frac{1}{8} \left(3 \delta^i_k 3^l 8^j_l-
         \delta^j_k 3^l 8^i_l\right)}_{3}.      
       \end{eqnarray}
\item The tensor decomposition of the operator $(b\bar{u})(u\bar{d})$ is found
 in
      Eq. \ref{b3x8}. Including a linear SU(3) breaking term, $(s\bar{s})$,
we have $(\overline{15}_{(1)}\oplus 6 \oplus \bar{3}) \otimes 
      (8 \oplus 1)$. Below we provide $\overline{15}\otimes 8$.
The others are already given. $6 \otimes 8$ can be obtained from
Eqn.~\ref{6x8} by moving all down indices up and all up indices down.
Below, ${\cal P}$ includes permutations of
all free upper or lower indices {\sl within} the parenthesis in which
it is found, but does not include permutations over the (symmetric)
lower two indices of the $\overline{15}$.  
      \begin{eqnarray}
%      \overline{15}^i_{jk} 8^l_m &=& 
%      \frac{1}{6} \left(\overline{15}^i_{jk} 8^l_m  + {\cal P} -
%      \frac{1}{6} \left\{ \delta^i_m \overline{15}^o_{jk} 8^l_o + 
%                          \delta^l_j \overline{15}^i_{ok} 8^o_m +
%                          {\cal P}\right\} \right.\nonumber\\
%      &&\underbrace{\left.+\frac{1}{30} \left\{ \delta^i_m \delta^l_j \overline{15}^o_{pk} 8^p_o + {\cal P}\right\}
%      \right)\qquad\qquad\qquad\qquad\qquad}_{\overline{42}}\nonumber\\
      \overline{15}^i_{jk} 8^l_m &=& 
     \underbrace{ \frac{1}{6} \left(\overline{15}^i_{jk} 8^l_m  + {\cal P} -
      \frac{1}{6} \left\{ \delta^i_m \overline{15}^o_{jk} 8^l_o + 
                          \delta^l_j \overline{15}^i_{ok} 8^o_m +
                          {\cal P}\right\}
      +\frac{1}{30} \left\{ \delta^i_m \delta^l_j 
       \overline{15}^o_{pk} 8^p_o + {\cal P}\right\}
      \right)}_{\overline{42}}\nonumber\\
      &+& \underbrace{\frac{1}{18} \epsilon_{jmo} \left( \overline{15}^i_{qk} 8^l_p \epsilon^{qpo} 
                   +{\cal P} - 
                   \frac{1}{5} \left\{ \delta^i_k \overline{15}^l_{qr} 8^r_p \epsilon^{qpo} 
                   +{\cal P} \right\} \right) + k\leftrightarrow j}_{24}\nonumber\\
%      &+& 
&&\hspace{-11mm}+\epsilon^{ilo} \underbrace{\frac{1}{24} 
      \left( \overline{15}^p_{jk} 8^q_m \epsilon_{pqo} + {\cal P}
      \right)}_{\overline{15'}}\nonumber\\
%      &+& 
&&\hspace{-11mm}+\underbrace{37 \delta^i_m {\overline{15}_{(3)}}^l_{jk} -
                      11 \delta^i_j {\overline{15}_{(3)}}^l_{mk} -
                      11 \delta^i_k {\overline{15}_{(3)}}^l_{jm} -
                      17 \delta^l_m {\overline{15}_{(3)}}^i_{jk} +
                      7  \delta^l_j {\overline{15}_{(3)}}^i_{mk} +
                      7  \delta^l_k {\overline{15}_{(3)}}^i_{jm}}_{\overline{15}_{(a)}}
      \nonumber\\
%      &+& 
&&\hspace{-11mm}+\underbrace{13 \delta^l_j {\overline{15}_{(4)}}^i_{km} +
                      13 \delta^l_k {\overline{15}_{(4)}}^i_{jm} -
                      11 \delta^l_m {\overline{15}_{(4)}}^i_{kj} -
                      5  \delta^i_j {\overline{15}_{(4)}}^l_{km} -
                      5  \delta^i_k {\overline{15}_{(4)}}^l_{jm} +
                      7  \delta^i_m {\overline{15}_{(4)}}^l_{kj}}_{\overline{15}_{(b)}}
      \nonumber\\
%      &+& 
&&\hspace{-11mm}+\underbrace{\frac{1}{4} \epsilon_{mko} \delta^i_j {6_{(2)}}^{lo} +
                      \frac{1}{4} \epsilon_{mjo} \delta^i_k {6_{(2)}}^{lo} -
                      \epsilon_{mjo} \delta^l_k {6_{(2)}}^{io} -
                      \epsilon_{mko} \delta^l_j {6_{(2)}}^{io}}_{6_{(a)}}\nonumber\\
%      &+& 
&&\hspace{-11mm}+\frac{1}{10} \delta^i_m \left(\delta^l_j \overline{15}^o_{kp} 8^p_o 
                                                     + k\leftrightarrow j\right)-
                       \frac{1}{40} \delta^i_j \left(\delta^l_k \overline{15}^o_{mp} 8^p_o 
                                                     + k\leftrightarrow m\right)\nonumber\\
      &&\hspace{-11mm} \underbrace{-\frac{1}{40} \delta^i_k \left(\delta^l_j \overline{15}^o_{mp} 8^p_o 
                      + j\leftrightarrow m\right),\qquad\qquad\qquad\qquad\qquad
         \quad}_{\bar{3}}
      \end{eqnarray}
      where we have used the following abbreviations
      \begin{eqnarray}
      {\overline{15}_{(3)}}^l_{jk} &=& 
           \frac{1}{72} \left(\overline{15}^o_{jk} 8^l_o - 
           \frac{1}{4} \left\{ \delta^l_j \overline{15}^o_{pk} 8^p_o +
                               \delta^l_k \overline{15}^o_{jp} 8^p_o\right\}\right),
      \nonumber\\
      {\overline{15}_{(4)}}^i_{km} &=&
           \frac{1}{72} \left(\overline{15}^i_{ok} 8^o_m + \overline{15}^i_{om} 8^o_k - 
           \frac{1}{4} \left\{ \delta^i_m \overline{15}^p_{ok} 8^o_p +
                              \delta^i_k \overline{15}^p_{om} 8^o_p\right\}\right),
      \nonumber\\
      {6_{(2)}}^{io} &=& \frac{1}{15} \left(
                         \overline{15}^i_{pq} 8^p_r \epsilon^{qro} +
                         \overline{15}^o_{pq} 8^p_r \epsilon^{qri}\right).
      \end{eqnarray}
      %
%      The tensor  decomposition of the $6 \otimes 8$ yields (here again
%${\cal P}$ does not include symmetrizing the indices on the 6)
%      %
%      \begin{eqnarray}
%      6^{ij} 8^l_m &=& \underbrace{\frac{1}{3} \left(6^{ij} 8^l_m 
%      + {\cal P} -\frac{1}{5} \left\{ \delta^i_m 6^{nj} 8^l_n  + {\cal P}\right\}
%      \right)}_{24}\nonumber\\
%      &-&\underbrace{\frac{1}{6} \epsilon^{ilo} \left( 
%      6^{qj} 8^p_m \epsilon_{opq} + {\cal P} - \frac{1}{4} \left\{
%      \delta^j_o 6^{qr} 8^p_m \epsilon_{rpq} + {\cal P}\right\}\right)+
%      i\leftrightarrow j}_{\overline{15}}\nonumber\\
%      &+&\underbrace{\frac{3}{20} \delta^i_m \left( 6^{jn} 8^l_n +
%      6^{ln} 8^j_n\right)+\frac{3}{20} \delta^j_m \left( 6^{in} 8^l_n +
%      6^{ln} 8^i_n\right)-\frac{1}{10} \delta^l_m \left( 6^{in} 8^j_n +
%      6^{jn} 8^i_n\right)}_{6}\nonumber\\
%      &-&\underbrace{\frac{1}{8} \left(\delta^i_m \epsilon^{jlo} +
%      \delta^j_m \epsilon^{ilo}\right) 6^{qr} 8^p_q
%      \epsilon_{pro}}_{\bar{3}}.
%      \end{eqnarray}
      %
\end{enumerate}

%********************************************************************************
\section{Complete $SU(3)$ flavor decomposition}
\label{completedecomp}
%********************************************************************************

For purposes of creating a self contained article we describe
briefly the method of the complete $SU(3)$ flavor decomposition, from
Ref.~\cite{gl}.
The physical amplitudes are decomposed into reduced amplitudes through
\begin{eqnarray} \label{SU3decomp}
\nonumber
{\cal A} (i^{R_c}_{\nu_c} \rightarrow f^{R_a}_{\nu_a}
f^{R_b}_{\nu_b}) &\!\!=\!\!&  (-1)^{\left(I_3+\frac{Y}{2} + \frac{T}{3}
\right)_{\overline{R}_c}}
\sum_{\stackrel{\scriptstyle R', \nu'}{R, \nu}}
\underbrace{\left( \begin{array}{ccc} 
                R_a & R_b & R' \\ \nu_a & \nu_b & \nu'
            \end{array} \right)}_
           {{\boldmath R_a}\otimes
            {\boldmath R_b}\rightarrow {\boldmath R'}}
\underbrace{\left( \begin{array}{ccc}
                R' & \overline{R}_c & R \\ \nu' & -\nu_c & \nu
            \end{array} \right)}_
           {{\boldmath R'}\otimes
            {\boldmath \overline{R}_c}\rightarrow {\boldmath R}}
\left< R' || R_\nu || R_c \right> .
\\
\end{eqnarray}
This convention ensures that the transformation matrix between the
physical and group theoretical basis is orthogonal. The notation for
the $SU(3)$ Clebsch-Gordan coefficients is from {\em{de Swart}}
\cite{deswart}. The representations of the initial state and the two
final states are named ${\boldmath R_c}$ and ${\boldmath R_a}$,
${\boldmath R_b}$, respectively; $\nu_i$ stands for the quantum
numbers $I_i,\,\,I_{3i},\,\,Y_i$; $T$ is the triality of the
representation ${\boldmath \overline{R}_c}$, defined as follows (using
Dynkin lables $n$ and $m$) (see, e.g., Ref.~\cite{georgi:group}):
\begin{equation}
T = (n-m) \,\, {\rm mod} \,\, 3 \stackrel{\rm for\,example}{=}
\left\{\begin{array}{cccc}
 1 & {\rm for} & {\boldmath 3} & (1,0), \\
-1 & {\rm for} & {\boldmath \overline{3}} & (0,1), \\
 0 & {\rm for} & {\boldmath 1} & (0,0), \\
 0 & {\rm for} & {\boldmath 8} & (1,1).
\end{array}
\right.
\end{equation}
The coefficient coupling the representations ${\boldmath R_a}\otimes
{\boldmath R_b}\rightarrow {\boldmath R_c}$ is given by \cite{gl}
\begin{eqnarray}
\left(
\begin{array}{ccc}
R_a   & R_b   & R_c\\[3mm]
\nu_a & \nu_b & \nu_c
\end{array}
\right)
&\equiv&
\left(
\begin{array}{ccc}
R_a   & R_b   & R_c\\[3mm]
I_a\,I_{a3}\,Y_a& I_b\,I_{b3}\,Y_b & I_c\,I_{c3}\,Y_c 
\end{array}
\right)
\nonumber\\[3mm]
&=&
\underbrace{F(R_c,\,Y_c,\,I_{c};\,R_a,\,Y_a,\,I_{a};\,R_b,\,Y_b,\,I_{b})}_
            {\rm isoscalar\,\, factor} \quad\times
 \underbrace{\langle I_c\,I_{c3} | I_a\,I_{a3}\,I_b\,I_{b3}\rangle}_
            {SU(2)\,\, {\rm Clebsch\,\, Gordan\,\, coefficient}}.
\end{eqnarray}
Tables of isoscalar factors are presented in \cite{kaeding}\footnote{Note the
following typo: The replacement
$\overline{24}\leftrightarrow 24$ has to be made 
in the tables 11, 16, 20, 35, 36 and 39 of Ref.~\cite{kaeding}}. 
The necessary SU(2) Clebsch Gordan 
coefficients are given for example in \cite{{deswart},{PDG}} or can be
easily computed using {\em{Mathematica}}. For the treatment of
final-state particles in the same representation or identical
particles in  the final state, see Ref.~\cite{gl}.\\[5mm]
Below we present the transformation from the physical to the 
group-theoretical basis (Eqn. \ref{SU3decomp}) for a class of
processes $in\rightarrow out_1 \ out_2$ as follows:
\begin{itemize}
\item Cabbibo allowed processes $a^{in}_{out_1,out_2} =
      A^{in}_{out_1,out_2} \cdot \alpha^{in}_{out_1,out_2}$,
\item Cabbibo suppressed processes $s^{in}_{out_1,out_2} =
      S^{in}_{out_1,out_2} \cdot \sigma^{in}_{out_1,out_2}$.
\end{itemize}

\input{appendix2a}

\newpage

\mbox{
\epsffile{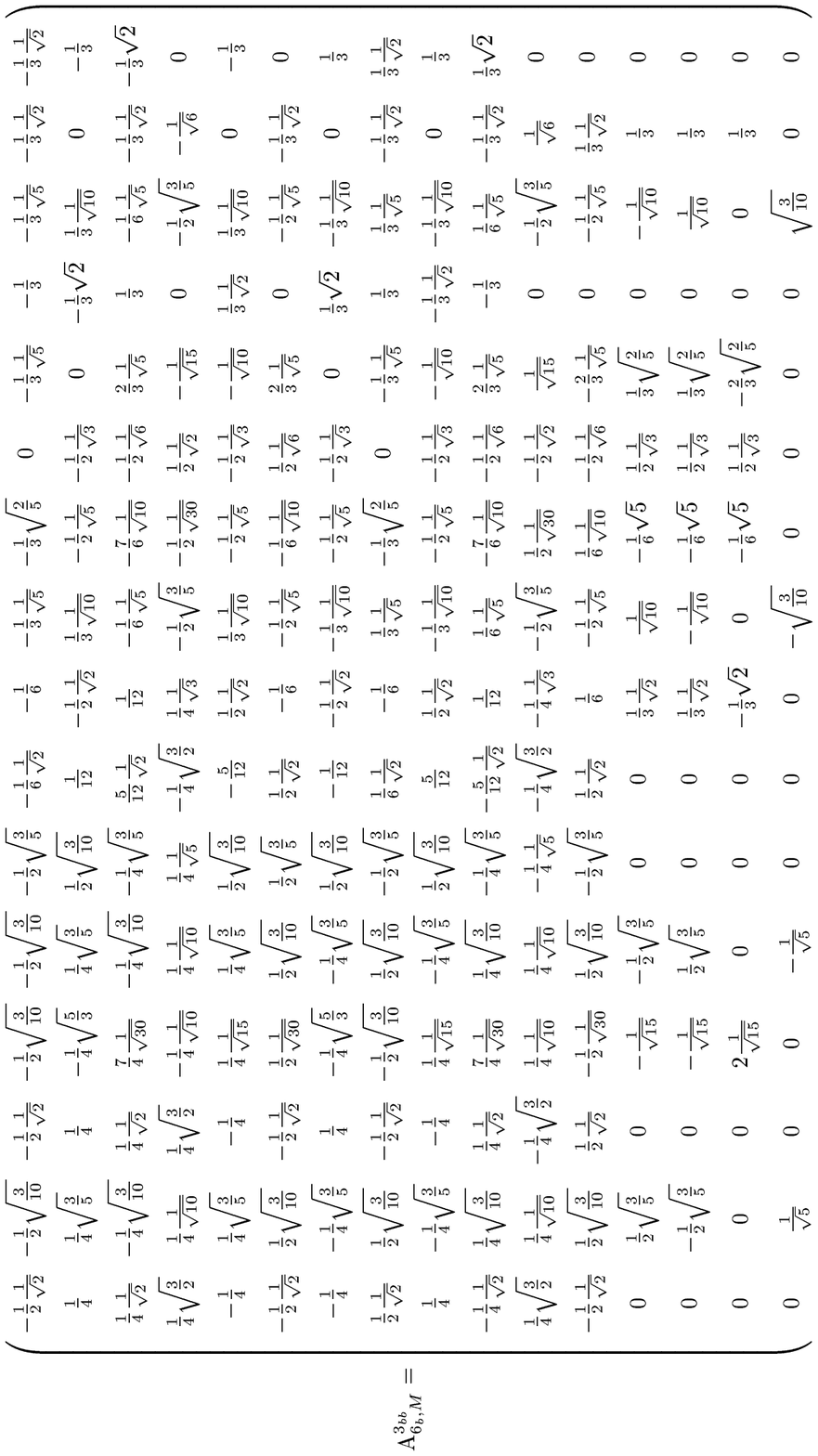}
\addtocounter{equation}{+1}
{(\Alph{section}\arabic{equation})}}

\mbox{
\epsffile{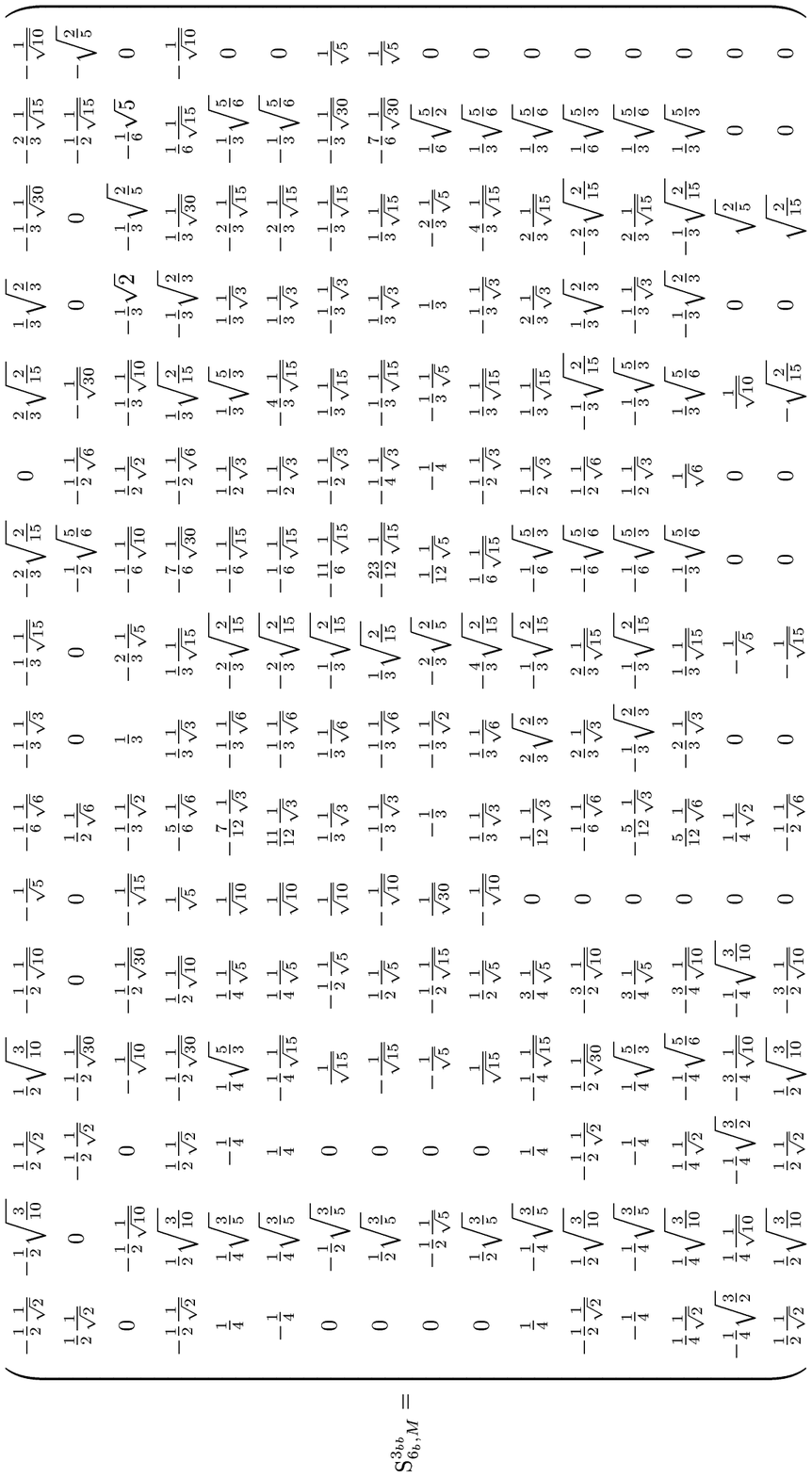}
\addtocounter{equation}{+1}
{(\Alph{section}\arabic{equation})}}

\input{appendix2b}

%********************************************************************************
\section{Matching arbitrary SU(3) breaking onto linear SU(3) breaking}
\label{matching}
%********************************************************************************

In Sec. \ref{decamps} we presented SU(3) decompositions of various decays
using tensor methods and asuming linear SU(3) breaking. In Appendix
\ref{completedecomp} we gave a full SU(3) decomposition. In this section
we show how to match these results in order to find relations for the reduced 
amplitudes given in Appendix \ref{completedecomp}. We write 
the unknowns given in Sec. \ref{decamps} and the reduced amplitudes from 
Appendix \ref{completedecomp} as
$U_i = (A,\ B,\ \ldots,A_{\rm LB},\ B_{\rm LB}, \ldots)_i$ and
$R_i = (\langle R'||R_\nu||R_{\rm out}\rangle,\ \ldots)_i$, respectively. We
then have
\begin{eqnarray}
{\rm process}_i = T_{ij} \ U_j,\qquad i_{\rm max}\geq j_{\rm max},\nonumber\\
{\rm process}_i = S_{ij} \ R_j,\qquad i_{\rm max} =  j_{\rm max}.
\end{eqnarray}
The matrices $S$ and $T$ are the transformation matrices. $S$ is orthogonal.
After imposing the same phase convention for both methods (see \cite{gl}),
we require
\begin{equation}
S_{ij} \ R_j = T_{ik} \ U_k \ \ \Longrightarrow \ \ R_l = S_{il} \ T_{ik} \ U_k,
\end{equation}
which automatically produces relationships between different reduced amplitudes,
holding in the case of linear breaking. For the sign convention see
the discussion in Sec. II of Ref. \cite{gl}. The amplitude expressions
have an overall sign ambiguity.  In order to find a consistent 
matching a sign change in the physical amplitudes considered 
with the tensor approach is necessary every time one of the 
following particles 
appears: $\pi^0$, $\pi^-$, $\eta_8$, $K^-$, $D^0$, $B^-$, $\Lambda_b^0$, $\Sigma^0$,
$\Lambda^0$, $\Sigma^-$ and $\Xi^-$.

Note that an operator employed using the tensor methods described
in the body of this paper is expressed in terms of its barred
components in the arbitarily broken SU(3) reduced matrix elements
used in Appendix \ref{completedecomp}.

%************************************************************************
\section{Phase space corrections}
\label{phasespacecorr}
%************************************************************************

In those $b$-decays where the final-state masses are not small compared
to the energy release, we need to include phase space effects in
going from amplitude relationships to rate relationships. 
In $B$-meson decays these corrections are normally small, but for
some $b$ baryon decays we can potentially find significant effects if the 
angular momentum $l$ of the decay channel is different from zero. 
Here we present the equations needed to include phase space corrections.
As new particle masses are measured, the phase space ``SU(3) breaking
effects'' can be separated from the dynamical ones.

For two-body decays where the decaying particle is in the rest frame
we can use \cite{PDG}
\begin{equation}
d\Gamma(a\rightarrow b c) = \frac{1}{32\pi^2} \left| {\cal M}
(a\rightarrow b c)\right|^2 \frac{|\vec{p}_b|}{m_a^2} d\Omega,
\end{equation}
where $|\vec{p}_b|=|\vec{p}_c|$ and 
\begin{equation}
|\vec{p}_b|=\frac{\left[(m_a^2-(m_b+m_c)^2)(m_a^2-(m_b-m_c)^2)\right]^{l+1/2}}
                 {2 m_a}.
\end{equation}
The above equation leads to the following phase space correction factor 
\cite{cbaryon}
\begin{equation}
\frac{\Gamma_l(a\rightarrow b c)}{\Gamma_l(d\rightarrow e f)}
= \left[\frac{\left[1-\displaystyle{\left(\frac{m_b+m_c}{m_a}\right)^2}\right]
              \left[1-\displaystyle{\left(\frac{m_b-m_c}{m_a}\right)^2}\right]}
             {\left[1-\displaystyle{\left(\frac{m_e+m_f}{m_d}\right)^2}\right]
              \left[1-\displaystyle{\left(\frac{m_e-m_f}{m_d}\right)^2}\right]}
        \right]^{l+1/2}
 \frac{\left| {\cal M}(a\rightarrow b c)\right|^2}
      {\left| {\cal M}(d\rightarrow e f)\right|^2}.
\end{equation}
We use Particle Data Book \cite{PDG} masses where they are measured, and
theoretical calculations \cite{spec,workshop,workspec} for the masses
of doubly heavy baryons. 

As an example, consider
Eqn.~\ref{bcMrate}.  
From Ref.~\cite{workshop,ccc} we have $m_{\Xi_{bb}^0} \approx
m_{\Xi_{bb}^-} \approx 10235 \,{\rm MeV}$, $m_{\Omega_{bb}^-} \approx
10385 \,{\rm MeV}$, $m_{\Xi_{bc}^+} \approx m_{\Xi_{bc}^0} \approx
6938 \,{\rm MeV}$ and $m_{\Omega_{bc}^0} \approx 7095 \,{\rm MeV}$.
This gives the following corrected relationships:
\begin{equation}
\frac{\Gamma\left(\Xi^0_{bb} \rightarrow \Xi^{+}_{bc} \ \pi^{-}\right)}
     {\Gamma\left(\Xi^0_{bb} \rightarrow \Xi^{+}_{bc} \ K^{-}\right)}
={1 \over \lambda^2} \ (1.02)^{l+1/2}, \qquad\qquad
\frac{\Gamma\left(\Xi^{-}_{bb} \rightarrow \Xi^0_{bc} \ \pi^{-}\right)}
     {\Gamma\left(\Omega^{-}_{bb} \rightarrow \Omega^{0}_{bc} \ K^{-}\right)}
= {1 \over \lambda^2} \ (1.01)^{l+1/2}.
\end{equation}
These are not significant corrections.  We might expect larger corrections
for final states containing a $b$ and {\sl two} charm quarks.
In Eqn.~\ref{bcbarDrate}, the relationships are from isospin and so
we do not expect corrections.  However, consider relationships from
Eqn.~\ref{6bjpsirate}, where we use the estimate of masses of the $6_b$ by
supposing that the splitting between $\Sigma_b$ and $\Lambda_b$,
for instance, tracks
that of the splitting between the $\Sigma_c$ and the $\Lambda_c$ 
\cite{bbaryons}:
\begin{equation}
\frac{\Gamma \left(\Xi^0_{bb} \rightarrow \Xi^0_{b2} \ J/\Psi \right)}
     {\Gamma \left(\Omega^-_{bb} \rightarrow \Omega^-_{b} \ J/\Psi \right)}
  = \frac{1}{2} \ (0.98)^{l+1/2}, \qquad\qquad
\frac{\Gamma \left(\Xi^-_{bb} \rightarrow \Sigma^-_{b} \ J/\Psi \right)}
   {\Gamma \left(\Omega^-_{bb} \rightarrow \Xi^-_{b2} \ J/\Psi \right)}
= 2 \ (0.98)^{l+1/2}.
\end{equation}
For more than two particles in the external state
the situation becomes harder to analyze. However, to the level we
are working (expanding about perfect SU(3) symmetry), we do not
expect phase space issues to be a significant source of error.  Only
if we include mass splittings on the order of 300 MeV for both isospin
breaking and strange quark mass differences do we begin to see
large phase space corrections.

\end{appendix}

\end{document}

%% file: SU3exacta.tex
{\footnotesize{
\begin{center}

\end{center}
}}

%% file: SU3exactb.tex
{\footnotesize{
\begin{center}
%
\end{center}
}}

%% file: pro3bb6bM.tex
{\footnotesize

\begin{center}
\begin{tabular}{|c|ccc|}
\hline
\hline
\TT $\,A-2\,\DLB-2\,\ELB$&$\XbbO$&$\rightarrow$&$\Sbp\Km$\\
\hline
\TT ${\frac{1}{{\sqrt{2}}}}\,A-{\sqrt{2}}\,\DLB-{\sqrt{2}}\,\ELB$&$\XbbO$&$\rightarrow$&$\SbO\barKO$\\
\hline
\TT ${\frac{1}{2}}\,B-\,\ELB-2\,\FLB$&$\XbbO$&$\rightarrow$&$\XbZO\PO$\\
\hline
\TT $-{\frac{1}{{\sqrt{3}}}}\,A+{\frac{1}{2}}{\frac{1}{{\sqrt{3}}}}\,B-2{\sqrt{3}}\,\DLB+{\frac{1}{{\sqrt{3}}}}\,\ELB-2{\frac{1}{{\sqrt{3}}}}\,\FLB$&$\XbbO$&$\rightarrow$&$\XbZO\EA$\\
\hline
\TT ${\frac{1}{{\sqrt{2}}}}\,B-{\sqrt{2}}\,\ELB-2{\sqrt{2}}\,\FLB$&$\XbbO$&$\rightarrow$&$\XbZm\Pp$\\
\hline
\TT $\,B-2\,\ELB+4\,\FLB$&$\XbbO$&$\rightarrow$&$\Obm\Kp$\\
\hline
\TT ${\frac{1}{{\sqrt{2}}}}\,A-{\sqrt{2}}\,\DLB-{\sqrt{2}}\,\ELB$&$\Xbbm$&$\rightarrow$&$\SbO\Km$\\
\hline
\TT $\,A-2\,\DLB-2\,\ELB$&$\Xbbm$&$\rightarrow$&$\Sbm\barKO$\\
\hline
\end{tabular}
\end{center}
\newpage

\begin{center}
\begin{tabular}{|c|ccc|}
\hline
\TT $\lambda \, \big(-{\frac{1}{{\sqrt{2}}}}\,B+{\frac{1}{{\sqrt{2}}}}\,\rho\,\DD-{\frac{1}{{\sqrt{2}}}}\,\rho\,F+3{\frac{1}{{\sqrt{2}}}}\,\rho\,H-{\sqrt{2}}\,\rho\,\II-{\frac{1}{{\sqrt{2}}}}\,\rho\,\BLB+3{\sqrt{2}}\,\rho\,\CLB$&$\XbbO$&$\rightarrow$&$\XbZm\Kp$\\
\TT $+{\frac{1}{{\sqrt{2}}}}\,\ELB-3{\sqrt{2}}\,\FLB-{\sqrt{2}}\,\rho\,\GLB-2{\sqrt{2}}\,\rho\,\HLB-{\sqrt{2}}\,\rho\,\ILB-11{\sqrt{2}}\,\rho\,\JLB\big)$&&&\\\hline
\TT $\lambda \, \big(-{\frac{1}{{\sqrt{2}}}}\,A-{\frac{1}{{\sqrt{2}}}}\,B+{\frac{1}{{\sqrt{2}}}}\,\rho\,C+{\frac{1}{{\sqrt{2}}}}\,\rho\,\DD+{\frac{1}{{\sqrt{2}}}}\,\rho\,\EE+3{\frac{1}{{\sqrt{2}}}}\,\rho\,G-{\sqrt{2}}\,\rho\,H$&$\Xbbm$&$\rightarrow$&$\SbO\Pm$\\
\TT $+3{\sqrt{2}}\,\rho\,\II-{\frac{1}{{\sqrt{2}}}}\,\rho\,\ALB-{\sqrt{2}}\,\rho\,\BLB-{\sqrt{2}}\,\rho\,\CLB+{\frac{1}{{\sqrt{2}}}}\,\DLB+{\sqrt{2}}\,\ELB+{\sqrt{2}}\,\FLB$&&&\\\TT $-{\sqrt{2}}\,\rho\,\GLB-{\frac{1}{{\sqrt{2}}}}\,\rho\,\HLB-3{\frac{1}{{\sqrt{2}}}}\,\rho\,\ILB+3{\sqrt{2}}\,\rho\,\JLB\big)$&&&\\\hline
\TT $\lambda \, \big({\frac{1}{{\sqrt{2}}}}\,A+{\frac{1}{{\sqrt{2}}}}\,B-{\frac{1}{{\sqrt{2}}}}\,\rho\,C-{\frac{1}{{\sqrt{2}}}}\,\rho\,\DD-{\frac{1}{{\sqrt{2}}}}\,\rho\,\EE+5{\frac{1}{{\sqrt{2}}}}\,\rho\,G+{\sqrt{2}}\,\rho\,H$&$\Xbbm$&$\rightarrow$&$\Sbm\PO$\\
\TT $+{\sqrt{2}}\,\rho\,\II+{\frac{1}{{\sqrt{2}}}}\,\rho\,\ALB+{\sqrt{2}}\,\rho\,\BLB+{\sqrt{2}}\,\rho\,\CLB-{\frac{1}{{\sqrt{2}}}}\,\DLB-{\sqrt{2}}\,\ELB-{\sqrt{2}}\,\FLB$&&&\\\TT $-{\frac{1}{{\sqrt{2}}}}\,\rho\,\GLB+{\frac{1}{{\sqrt{2}}}}\,\rho\,\HLB+3{\frac{1}{{\sqrt{2}}}}\,\rho\,\ILB+9{\frac{1}{{\sqrt{2}}}}\,\rho\,\JLB\big)$&&&\\
\hline
  \hline
\end{tabular}
\end{center}
}
%%% Local Variables: 
%%% mode: latex
%%% TeX-master: "pro3bb3bcMM"
%%% End: 

%% file: appendix2a.tex
\subsection{$3_{bb}\longrightarrow 3_{bc}+M$}
\label{3bb3bcM}

\begin{equation}

\right)
.\end{equation}

\subsection{$3_{bb}\longrightarrow 6_b+D$}
\label{3bb6bD}

\begin{equation}
%
\right)
.\end{equation}

\subsection{$3_{bb}\longrightarrow \bar{3}_b+D$}
\label{3bbbar3bD}

\begin{equation}
%
\end{equation}

\subsection{$3_{bb}\longrightarrow 3_{bc}+\overline D$}
\label{3bb3bcbarD}

\begin{equation}
%
\right)
.\end{equation}

\subsection{$3_{bb}\longrightarrow \bar{3}_b+J/\Psi$}
\label{3bbbar3bJP}

\begin{equation}
%
\end{equation}

\subsection{$3_{bb}\longrightarrow 6_b+J/\Psi$}
\label{3bb6bJP}

\begin{equation}
%
\end{equation}

\subsection{$3_{bb}\longrightarrow \bar{3}_b+M$}
\label{3bbbar3bM}

\begin{equation}
%
\right)
.\end{equation}

\subsection{$3_{bb}\longrightarrow 6_b+M$}
\label{3bb6bM}

\begin{equation}
%
\end{equation}

%%% Local Variables: 
%%% mode: latex
%%% TeX-master: t
%%% End: 

%% file: appendix2b.tex
\subsection{$3_{bb}\longrightarrow B+b$}
\label{3bbBb}

\begin{equation}
%
\right)
.\end{equation}

\subsection{$3_{bb}\longrightarrow \bar{3}_b+\overline D$}
\label{3bbbar3bbarD}

\begin{equation}
%
\right)
.\end{equation}

\subsection{$3_{bb}\longrightarrow 6_b+\overline D$}
\label{3bb6bbarD}

\begin{equation}
%
\right)
.\end{equation}

%%% Local Variables: 
%%% mode: latex
%%% TeX-master: t
%%% End: 

%% file: paper.bbl
\begin{references}
\nopagebreak

\bibitem{quiggtalk} C. Quigg, ``Doubly Heavy Baryons,'' in Workshop on
B Physics at the Tevatron: RunII and Beyond, Sept. 25, 1999.

\bibitem{NRQCD} W.E. Caswell and G.P. Lepage, Phys. Lett. {\bf B167} (1986)
437; G.T. Bodwin, E. Braaten, and G.P. Lepage, Phys. Rev. {\bf D51} (1995)
1125; Erratum-ibid. {\bf D55} (1997) 5853.

\bibitem{HQET} N. Isgur and M.B. Wise, Phys. Lett. {\bf B232} (1989) 113;
Phys. Lett. {\bf B237} (1990) 527; H. Georgi, Phys. Lett. 
{\bf B240} (1990) 447;
E. Eichten and B. Hill, Phys. Lett. {\bf B243} (1990) 427.

\bibitem{sumrules}
V.A. Novikov, L.B. Okun, M.A. Shifman, A.I. Vainshtein, M.B. Voloshin, V.I.
Zakharov, Phys. Rept. {\bf 41} (1978) 1; 
M.A. Shifman, 
A.I. Vainshtein, and V.I. Zakharov, Nucl. Phys. {\bf B147} (1979) 385;
L.J. Reinders, H.R. Rubinstein, S. Yazaki, Phys. Rept. {\bf 127} (1985) 1;
M.B. Voloshin, Int.
J. Mod. Phys. {\bf A10} (1995) 2865.

\bibitem{spec} 
E. Bagan, M. Chabab, S. Narison, Phys. Lett. {\bf B306} (1993) 350;
%baryons with two heavy quarks from qcd spectral sum rules
E. Bagan, H.G. Dosch, P. Gosdzinsky, S. Narison, J.M. Richard, 
  Z. Phys. {\bf C64} (1994) 57;
%Hadrons with Charm and Beauty
S.S. Gershtein, V.V. Kiselev, A.K. Likhoded, A.I. Onishchenko,
Phys. Rev. {\bf D62} (2000) 054021;
% spectroscopy of doubly heavy baryons.
V.V. Kiselev and A.I. Onishchenko, Nucl. Phys. {\bf B581} (2000) 432;
%doubly heavy baryons in sum rules of NRQCD
V.V. Kiselev, A.K. Likhoded, O.N. Pakhomova, and V.A. Saleev,
Phys. Rev. {\bf D66} (2002) 034030.
%mass spectra of doubly heavy omega baryons.

\bibitem{bc} F. Abe et al., Phys. Rev. Lett. {\bf 81} (1998) 2432;
Phys. Rev. {\bf D58} (1998) 112004.

\bibitem{cchadron} M. Mattson et al., Phys. Rev. Lett. {\bf 89} (2002) 112001.
%SELEX collaboration, FNAL, first obs. of xi^+(cc)


\bibitem{workshop} K. Anikeev et al.,
``B Physics at the Tevatron: Run II and Beyond,''
hep-ph/0201071.

\bibitem{workspec} 
S. Fleck and J.M. Richard, Prog. Theor. Phys. {\bf 82} (1989) 760;
M.J. Savage and M.B. Wise, Phys. Lett. {\bf B248} (1990) 177;
%spectrum of baryons with two heavy quarks
J. Richard, hep-ph/9407224; 
R. Roncaglia, D.B. Lichtenberg, 
E. Predazzi, Phys. Rev. {\bf D52} (1995) 1722; 
A. Martin, J.-M. Richard, Phys. Lett. {\bf B355} (1995) 345;
%Omega_c and other charmed baryons revisited.
M.L. Stong, hep-ph/9505217;
% predicting masses of baryons containing one or two heavy quarks.
D.B. Lichtenbert, R. Roncaglia, and E. Predazzi, Phys. Rev. {\bf D53} 
   (1996) 6678;  
D.U. Matrasulov, M.M Musakhanov
   and T. Morii, Phys. Rev. {\bf C61} (2000) 045204; 
S.-P. Tong et al., Phys. Rev. {\bf D62} (2000) 054024;
%spectra of baryons containing two heavy quarks in pot. model
D.U. Matrasulov, F.C. Khanna, Kh.Yu. Rakhimov, and Kh.T. Butanov, Eur. Phys.
J. {\bf A14} (2002) 81;
%spectra of baryons containing two heavy quarks.
V.V. Kiselev and A.K. Likhoded, Phys. Usp. {\bf 45} (2002) 455; 
%baryons with two heavy quarks 109 pages.
D. Ebert, R.N. Faustov, V.O. Galkin, and A.P. Martynenko, Phys. Rev.
{\bf D66} (2002) 014008.
%mass spectra of doubly heavies in Rel QM


\bibitem{wkdecay} M.J. White and M.J. Savage, Phys. Lett. {\bf B271} 
  (1991) 410;
%semileptonic decays of baryons with two heavy quarks.
M.A. Sanchis-Lozano, Nucl. Phys. {\bf B440} (1995) 251;
%weak decays of doubly heavy hadrons.
%Phys. Lett. {\bf B321} (1994) 407; 
%semilep decays of baryons containing two heavy quarks.
X.-H. Guo, H.-Y. Jin, and X.-Q. Li, Phys. Rev. {\bf D58} (1998) 114007. 
%weak semileptonic decays of heavy baryons containing two heavy quarks.
\bibitem{onish} A.I. Onishchenko, hep-ph/0006295; hep-ph/0006271.


\bibitem{quigg} M.B. Einhorn and C. Quigg, Phys. Rev. {\bf D12} (1975) 2015.
%nonlep decays of charmed mesons.
\bibitem{sw1} M.J. Savage and M.B. Wise, Phys. Rev. {\bf D39} (1989) 3346;
{\bf D40} (1989) 3127. 
% SU3 for nonlep B decays

\bibitem{Bmesons} 
D. Zeppenfeld, Z. Phys. {\bf C8} (1981) 77;
M.J. Savage and M.B. Wise, Nucl. Phys. {\bf B326} (1989) 15;
%SU3 pred for nonlep B -> charmed baryons
S.M. Sheikholeslami, G.K. Sidana, and M.P. Khanna,
Int. J. Mod. Phys. {\bf A7}, (1992) 1111.
% SU3 B meson to baryons.



\bibitem{phases}
M. Gronau, O.F. Hernandez, D. London, and J.L. Rosner,
Phys. Rev. {\bf D50} (1994) 4529; % B mesons to two light pseudoscalars
A.J. Buras and R. Fleischer, Phys. Lett. {\bf B341} (1995) 379;
%Limitations in measuring the angle beta using SU(3) relations
A.S. Dighe, M. Gronau, and J.L. Rosner, Phys. Lett. {\bf B367} (1996) 357;
Erratum-ibid. {\bf B377} (1996) 325; 
% amplitude relations B -> etas; phases and some breaking.
X.-G. He, Eur. Phys. J. {\bf C9} (1999) 443; % su3 of annhil and CP viol.
% for B -> PP.
N.G. Deshpande, X.-G. He, J.-Q. Shi, Phys. Rev. {\bf D62} (2000) 034018.
%SU3 sym and CP viol B-> PV

%\bibitem{kingsley} R.L. Kingsley, S.B. Treiman, F. Wilczek, and A. Zee,
%Phys. Rev. {\bf D1} (1975) 1919.
% weak decays of charmed hadrons

\bibitem{bbaryons} R. Arora, 
  G.K. Sidana, and M.P. Khanna, Phys. Rev. {\bf D45},
(1992) 4121. % SU3 pred for nonlep weak of (single) bottom baryons.

\bibitem{ccbaryon} 
M.J. Savage and R.P. Springer, Int. J. Mod. Phys. {\bf A6}, 
(1991) 1701.  %very charming baryons.

\bibitem{gl} B. Grinstein and R.F. Lebed, Phys. Rev. {\bf D53}
(1996) 6344.


\bibitem{du}
D.-S. Du and D.-X. Zhang, Phys. Rev. {\bf D50} (1994) 2058.
%SU3 breaking in nonlep of bottom baryons.



\bibitem{Bbreak} 
M. Gronau, O.F. Hernandez, D. London, and J.L. Rosner,
Phys. Rev. {\bf D52} (1995) 6356;
% broken SU(3) in 2 body B decays
H.J. Lipkin, Phys. Lett. {\bf B415} (1997) 186;
% penguins, tree, % and FSI in B Decay in Broken SU3.
S. Oh, Phys. Rev. {\bf D60} (1999) 034006; % SU3 symm. and factorization
% to two charmless.  matches them.
M. Gronau and D. Pirjol, Phys. Rev. {\bf D62} (2000) 077301.
%extrap SU3 break from D to B decays


\bibitem{savage} M. Savage, Phys. Lett. {\bf B257} (1991) 414.
%SU3 viol of nonlep decay of charmed hadrons.

\bibitem{lightbreak}
W. Kwong and S.P. Rosen, Phys. Lett. {\bf B298} (1993) 413;
% minimal breaking of flavor SU3 in nonlep charm decay
I. Hinchliffe and T.A. Kaeding, Phys. Rev. {\bf D54} (1996) 014.
% nonlep two body decay of D in broken SU(3).

\bibitem{kiselev}
V.V. Kiselev and A.E. Kovalsky, Phys. Rev. {\bf D64} (2001) 014002.
%doubly heavy baryons omega vs. xi in sum rules of NRQCD





\bibitem{hg} H. Georgi, ``Weak Interactions and Modern Particle Theory,''
Benjamin/Cummings, 1984.



\bibitem{wolfram} Stephen Wolfram, ``The Mathematica Book'',
Cambridge University Press, 1999.

\bibitem{kaeding} T.A. Kaeding, ``Tables of SU(3) Isoscalar
Factors'', nucl-th/9502037.

\bibitem{wolf} L. Wolfenstein, Phys. Rev. Lett. 51 (1983) 1945.

\bibitem{georgi:group} H. Georgi, ``Lie Algebras in Particle Physics,''
Perseus Books, 1999.

\bibitem{deswart} J. J. de Swart, Rev. Mod. Phys. {\bf 35} (1963) 916;
                  Rev. Mod. Phys. {\bf 37} (1965) 326.

\bibitem{PDG} Particle Data Group, D.~E.~Groom et al., Eur. Phys. J. {\bf C15}
              (2000) 1.

\bibitem{cbaryon} M.J. Savage and R.P. Springer, Phys. Rev. {\bf D42}
(1990) 1527. % SU3 pred for charmed baryon decays.

\bibitem{ccc} J.D. Bjorken, ``Is the $ccc$ a new deal for baryon
spectroscopy?'' Fermilab-Conf-85/69.

\end{references}
